\documentstyle[emulateapj,apjfonts,epsfig]{article}

\begin{document}
\tightenlines

\renewcommand{\textfraction}{0.05}

\def\et{{et\,al.}}
\def\rd{Di\thinspace Stefano}
\def\etal{{et\,al.}}
\def\fdg{\hbox{$.\!\!^\circ$}}          % Fractions of degrees
\def\amin{\ifmmode ^{\prime}\else$^{\prime}$\fi}
\def\degs{\ifmmode ^{\circ}\else$^{\circ}$\fi}
\def\gc{globular cluster}
\def\rd{Di\thinspace Stefano}
\def\bi{binary} \def\gl{gravitational lens} \def\pr{probability}
\def\refpar{\hangindent=3em\hangafter=1}
\def\reference{\refpar\noindent}
\def\apj{ApJ}
\def\aap{A\&A}
\def\araa{ARA\&A}
\def\mnras{MNRS}

%\pagestyle{myheadings}
%\markboth{X-Ray sources in M31 globular clusters}{X-Ray sources in M31 globular clusters}
%\begin{document}

%\documentclass{article}
%\usepackage{epsfig,rotating,aaspp4}
\newcommand{\bol}{Bo\,375}
\newcommand{\chandra}{{\it Chandra}}
\newcommand{\einstein}{{\it Einstein}}
\newcommand{\rosat}{{\it ROSAT}}
\newcommand{\asca}{{\it ASCA}}
\newcommand{\xrs}{X-ray source}
\newcommand{\lus}{luminous}
\newcommand{\luf}{luminosity function} 
\newcommand{\lum}{\thinspace\hbox{$\hbox{erg}\thinspace\hbox{s}^{-1}$}}
\newcommand{\flux}{\thinspace\hbox{$\hbox{erg}\thinspace\hbox{cm}^{-2}\thinspace
\hbox{s}^{-1}$}}
\newcommand{\msun}{\thinspace\hbox{$M_{\odot}$}}
\newcommand{\mdot} {\thinspace\hbox{$\dot M$}}

\submitted{Accepted for publication in the Astrophysical Journal}
\title{Bright X-Ray Sources in M31 Globular Clusters}
\author{R.~Di\,Stefano\altaffilmark{1}, A. K. H. Kong, M. R.~Garcia, P.~Barmby}
\affil{Harvard-Smithsonian Center for Astrophysics, 60
Garden Street, Cambridge, MA 02138}
\author{J.~Greiner}
\affil{Astrophysical Institute Potsdam, 14482 Potsdam, Germany}
\author{S.S.~Murray,  F.A.~Primini} 
\affil{Harvard-Smithsonian Center for Astrophysics, 60
Garden Street, Cambridge, MA 02138}

\altaffiltext{1}{also Department of Physics and Astronomy, Tufts
University, Medford, MA 02155}

\begin{abstract}

We have conducted {\it Chandra} observations of $\sim 2560$ square
arcmin ($\sim 131$ kpc$^2$) of M31, and find that the most luminous
X-ray sources in most of our fields are in \gc s. Of the $28$ \gc\
\xrs s
in our fields, $15$ are newly discovered.   
Approximately $1/3$ of all the sources
have $L_X$([0.5--7]\, keV)$~ > 10^{37}$ ergs s$^{-1}$; 
approximately $1/10$ of all the sources
have $L_X$([0.5--7]\, keV) close to or above $10^{38}$ ergs s$^{-1}$.
The most luminous  source, in the globular cluster Bo 375,
is consistently observed to have $L_X$ greater than $2\times 10^{38}$
ergs s$^{-1}$.

\noindent (1) We present data on the spectra and/or light curves of
the $5$ most luminous M31 \gc\ sources. 

\noindent (2)  
We explore possible explanations for the high X-ray luminosities 
of the brightest sources. These include that 
the \xrs s may be composites, the radiation we receive may be
beamed, metallicity effects could be at work, or the sources may
be accreting black holes.
We weigh each of these possibilities against the data.
In addition, we introduce a neutron star 
model in which mass transfer proceeds on the thermal
time scale of the donor star. 
Our model  
can produce luminosities of several times $10^{38}$ ergs s$^{-1}$,
and leads to a set of well-defined predictions.   

\noindent (3) We compute the X-ray luminosity 
function and the distribution of
counts in wavebands that span the range of energies to which {\it Chandra} is sensitive.
We find the peak X-ray luminosity is higher and that systems with
$L_X > 10^{37}$ erg s$^{-1}$ constitute a larger fraction of all GC sources than in
our Galaxy.
 
\noindent (4) We study the possible reasons for this difference between M31 and Galactic 
globular cluster X-ray sources 
and identify three promising explanations. 
\end{abstract}  


\section{Introduction}

\subsection{X-Ray Sources in Galactic and M31 Globular Clusters} 

\subsubsection{Galactic Globular Cluster X-Ray Sources}

X-ray studies of Milky Way (MW) \gc s (GCs) suggested that they be divided into
``bright" ($L_X > 10^{35}$ ergs s$^{-1}$) and ``dim" ($L_X < 10^{34}$ ergs s$^{-1}$)
sources (see e.g. Hertz \& Grindlay 1983; Verbunt et al. 1985; Deutsch
et al. 2000). The majority of dim sources are thought to be accreting
white dwarfs (Di\,Stefano \& Rappaport 1994; Hakala et al. 1997;
Grindlay et al. 2001). 
%with many new discoveries of sources with $L_X<10^{32}$
%ergs s$^{-1}$. 
{\it Chandra} observations with realistic exposure
times can only see evidence of
large populations of dim sources in M31, not individual sources. 
We can therefore draw comparisons only between the bright GC sources
in each galaxy.
Twelve MW GCs house bright X-ray sources; until recently,
no cluster was known to have more than a single bright source
\footnote{White \& Angelini (2001) recently used {\it Chandra} to resolve two distinct luminous
X-ray sources separated by $< 3''$ in the massive post-core-collapse
MW GC, M15.}.
There
is evidence, largely from X-ray bursts, that the bright MW GC
sources are
accreting neutron stars. There is no evidence that any MW GC harbors
an accreting black hole. 
Indeed, this apparent lack of
black holes in MW GC has been commented upon and explained as perhaps
due to large kick velocities obtained by BH binaries upon interactions
with other GC members (Sigurdsson \& Hernquist 1993; Kulkarni et al. 1993).
The most luminous MW GC X-ray source
(4U\,1820--30 in NGC\,6624) has $L_X \sim 1-5\times 10^{37}$ ergs s$^{-1}$ (see
Bloser et al. 2000 for a review).

\subsubsection{M31 Globular Cluster X-Ray Sources}

We have conducted {\it Chandra} observations of $\sim 2560$ square
arcmin ($\sim 131$ kpc$^2$) of M31, and find that the most luminous
X-ray sources in most of our fields are in \gc s. Of the $28$ \gc\
\xrs s
in our fields, $15$ are newly discovered.   
Approximately $1/3$ of all the sources
have $L_X$([0.5--7]\, keV)$~ > 10^{37}$ ergs s$^{-1}$; 
approximately $1/10$ of all the sources
have $L_X$([0.5--7]\, keV) close to or above $10^{38}$ ergs s$^{-1}$.
The most luminous  source, in the globular cluster Bo\,375,
is consistently observed to have $L_X$ greater than $2\times 10^{38}$
ergs s$^{-1}$.
% and $5\times 10^{38}$
%ergs s$^{-1}$. 
Given the sensitivity of the detectors we  used (ACIS-S and
ACIS-I), and our exposure  times ($15$ ksec and $8.8$ ksec, respectively), 
we could have
detected sources with luminosities as low as 
$1.5-5 \times 10^{35}$ ergs s$^{-1}$, and indeed have discovered some
sources only slightly brighter than these limits.
The large fraction of \gc\ sources with $L_X > 10^{37}$ ergs s$^{-1}$
stands in sharp contrast with the situation in the  Galaxy,
where, e.g., only one of the 11 bright sources studied in the
{\it ROSAT} all-sky survey had $L_X > 10^{37}$ ergs s$^{-1}$ (Verbunt et al. 1995).

Each of the following subsections ($\S 1.2-1.6$) provides an overview of
the corresponding section ($\S 2-6$) of the paper.  

\subsection{Bo\,375:
the highest-luminosity X-ray source in a globular cluster} 

While conducting a census of M31 X-ray sources, we found that 
the most luminous source in our census was associated with
the globular cluster Bo\,375. Our observations were performed 
using the  ACIS-S  detector aboard the {\it Chandra} X-Ray Observatory.  
The uncertainty in the position of the source is $~\sim 1'',$ comparable
to the uncertainty in the cluster position from the Bologona catalog (Battistini et al. 1997). 
Since the density of X-ray sources in the region around the source 
is not very high (there is  just one other source   with $L\sim 10^{36}$
erg s$^{-1}$ within $5'$), it is likely that the X-ray source
is physically associated with Bo\,375. The measured luminosity (between
$0.5$ and $2.4$ keV) of this
source is between $\sim 2 - 5\times 10^{38}$ ergs s$^{-1}$.  
In \S 2 we present 
new data
on the flux, spectrum, and light curve
of the X-ray source, Bo\,375. In addition,
we place these observations in
perspective by summarizing and, in some cases re-analyzing,
data from previous X-ray observations of the source.
Although the spectrum and/or light curve of Bo\,375 was 
studied in 2 earlier
papers (Supper {\etal} 1997; Irwin \& Bregman 1999), the source
luminosity is not quoted in the literature. We find, however,
that, like our own observation,
 all previous observations of the X-ray flux, which is
variable, are consistent with luminosities that range from 
$\sim 2 \times 10^{38}$ ergs s$^{-1}$ to
$\sim 5 \times 10^{38}$ ergs s$^{-1}$. That is,  
barring beaming effects, the luminosity is most  
likely super-Eddington for a $1.4\, M_\odot$ accretor.
This would make Bo\,375 an interesting source whatever its
location, but its position within or very near to a globular
cluster (GC) makes it even more special. This is because, as mentioned above,
 the maximum luminosity observed for a MW GC 
X-ray source is $4$ times smaller than the minimum luminosity
found for Bo\,375.
In \S 2 we also explore possible physical reasons for the observed high
luminosity of this source. 

\subsection{Other Bright M31 GC X-Ray Sources} 
 
The second and third most luminous X-ray sources in our sample are also 
associated with \gc s, and one of these sources also exhibits
X-ray luminosity close to $10^{38}$ ergs s$^{-1}$.  
The fields we were surveying are
shown in Figure 1, superposed on an optical image of M31.  
Because the existence of several luminous sources
 suggests a more general phenomenon, we have considered an additional
data set, collected during 1999 and 2000
 by the combination of {\it Chandra's} HRC, ACIS-I, and ACIS-S.
The combined data sets include $10$ \gc\ \xrs s with $L_X > 10^{37}$ ergs s$^{-1}$.
The spectra and light curves of $4$ of these sources are presented in \S 3. 
The light curves of these sources (and also of Bo\,375), exhibit enough structure
on short time scales that it is clear that at least one source in the 
cluster must have a luminosity that is a significant fraction
of the total luminosity we have observed.
 
\begin{figure*}
\begin{center}
\psfig{file=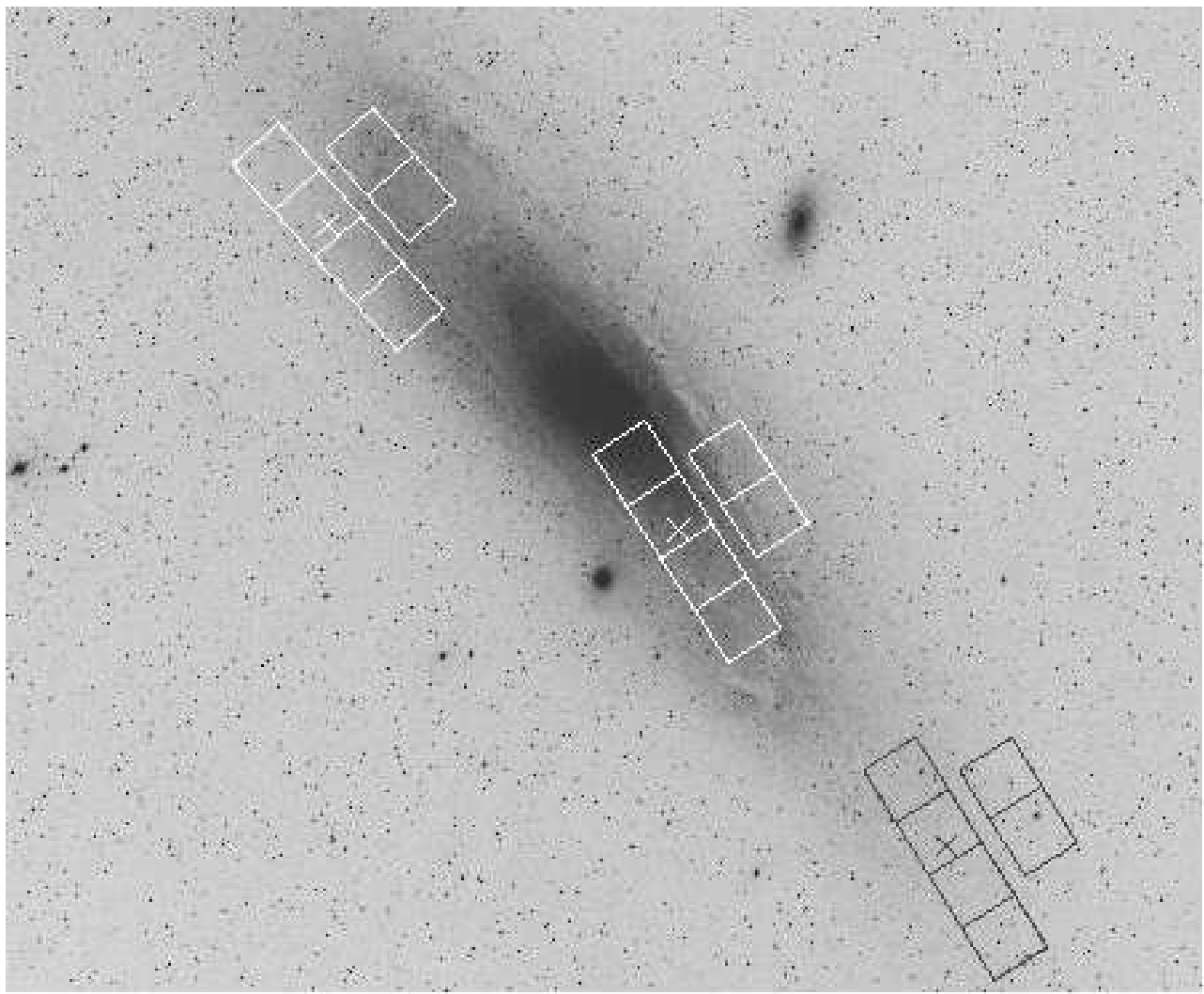,width=17cm}
\end{center}
\caption{The regions observed by the M31 AO2 GO census overlaid on
an optical Digitized Sky Survey image of M31. North is up, and
east is to the left. The chip orientation 
is shown for the $15$ ksec observations that occurred in November 2000. 
Subsequent observations have rotated by $90^{\circ}$ and then a 
further $120^{\circ}$. Each aim-point (always in S3) is marked by a
cross ($\times$).} 
\end{figure*}

 \subsection{Population Studies}
  
These high-luminosity sources raise questions about possible 
differences between the \luf\ of \gc\ \xrs s in M31 and that of the 
Milky Way. 
We therefore, in \S 4,
construct the luminosity function of the M31 \gc\ \xrs s as measured
by {\it Chandra}, and
compare it to the luminosity function of Milky Way \gc\ sources 
detected during the {\it ROSAT} all-sky survey (Verbunt \etal\ 1995).

\subsection{M31's Globular Clusters}

The primary differences between the population of X-ray sources in
M31 \gc s and those in Galactic \gc s are that (1) the peak
luminosity is higher, and (2) the high-luminosity end of the
distribution function is more populated.
It is natural to ask if these differences can be explained by differences in
the two galaxies' populations of \gc s.
This question is addressed in \S 5. 

\subsection{Conclusions: Possible Explanations} 

Explanations are required at two levels:
(1) what is the nature of the sources that appear to be so highly
luminous? (2) what properties of M31 and its system of  \gc s
generate these bright sources? 
These issues are the focus of \S 6.

\section{Bo\,375}

\subsection{Observations and Data Reduction}
\bol\ has been observed by several X-ray missions. A
summary of the X-ray observations 
used here is given in Table 1. \bol\ was first observed with
\einstein\ in 1979 during a survey of M31 and then it frequently
appeared in the \rosat\ M31 survey data from 1991 to 1994. It was also
observed with \asca\ in 1993 and \chandra\ in 2000 and 2001. We therefore have
observations spanning an interval of more than 20 years,
and can study the spectrum and the time evolution of the source.

\small{
\begin{table*}
\begin{center}
TABLE~1

{\sc Observation log of Bo\,375}

\vspace{1mm}
\begin{tabular}{l c c c c c}
\hline\hline
Date & Observatory & Instrument & Exposure & Count rate& Remarks \\
 & & &(ksec)&(c/s)&\\
\hline\\
7 August 1979 & \einstein\ & HRI & 9.9 & 0.03 &\\
25 July 1991 & \rosat\ & PSPC & 30.2  & 0.20 &\rosat\#1\\
26 July 1991 & \rosat\ & PSPC & 26.6  & 0.20 &\rosat\#2, near the `rib'\\
5 January 1991 & \rosat\ & PSPC & 44.7 & 0.11& \rosat\#3, near the `rib'\\
29 July 1993 & \asca\ & GIS & 32.4 & 0.09&\\
30 November 1999 & \chandra\ & HRC & 1.2  & 0.19&\\
13 Feburary 2000 & \chandra\ & HRC & 0.8  & 0.27&\\
8 March 2000 & \chandra\ & HRC & 2.1  & 0.25&\\
26 May 2000 & \chandra\ & HRC & 1.2 & 0.23&\\
21 June 2000 & \chandra\ & HRC & 1.2 & 0.31&\\
18 August 2000 & \chandra\ & HRC & 1.2  & 0.26&\\
11 September 2000 & \chandra\ & HRC & 1.2 & 0.21&\\
12 October 2000 & \chandra\ & HRC & 1.2 & 0.30&\\
1 November 2000 & \chandra\ & ACIS-S & 13.8  & 0.60&\\
1 Feburary 2001 & \chandra\ & HRC & 1.2 & 0.23&\\
\hline \\
\multicolumn{6}{l}{NOTES --- This list includes only those
observations we have used in this paper.} 
\end{tabular}
\end{center}
\end{table*}
}

\subsubsection{Previous X-ray Observations}

{\it Einstein Data} ---
\bol\ was observed with \einstein\ High Resolution Imager (HRI) in
1979 during M31 survey observations. The \einstein\ HRI has a spatial
resolution of $\sim 3''$, and no spectral resolution. The count rate
was extracted from a $18''$ radius circle centered on the source
centroid and the count rate is $\sim 0.03$ counts s$^{-1}$,
corresponding to $\sim 5\times 10^{38}$ ergs s$^{-1}$ in 0.5--10 keV
(assuming a power-law model with $N_H=10^{21}$ cm$^{-2}$, $\alpha=1.7$
and a distance of 780 kpc; Macri et al. 2001). 
For full details on source counts and
background extraction and events corrections we refer to Fabbiano (1988).

{\it ROSAT Data} ---
For the energy spectra of \bol, we only analyzed only long ($> 20$
ksec) observations. There are three long observations in the archival
database, and two of them (\rosat\#2 and \rosat\#3) are just outside the
`rib' support structure at $18'$ from the center of the field. The
remaining one (\rosat\#1) is $\sim 14'$ off-axis. Each spectrum was extracted from a circular
aperture centered on the source. The size of the extraction radius
varied from $7''$ to $9''$ with the off-axis angle of the source to
account for the point-spread function of the instrument. Background
was subtracted from an annulus centered on the source. The spectra
were binned so that there are at least 20 photons in each energy bin,
and all channels below 0.1 keV and above 2.4 keV were ignored.

{\it ASCA Data} ---
The {\it Advanced Satellite for Cosmology and Astrophysics} (\asca)
satellite (Tanaka \etal\ 1994) is equipped with two Solid State
Imaging Spectrometers (SIS; 0.4--10 keV) and two Gas Imaging
Spectrometers (GIS; 0.7--10 keV). For our purpose, we only used the
data taken by GIS instruments, because the source was near the edge of
SIS. \asca\ observed \bol\ on 1993 July 29. Standard data screening
was employed \footnote{See the \asca\ Data Reduction Guide 2.0
(http://heasarc.gsfc.nasa.gov/docs/asca/abc/abc.html)}. Data taken at
a geomagnetic cut-off rigidity lower than 4 GeV, at an elevation angle
less than $5^\circ$ from the Earth's limb, and during passage through the
South Atlantic Anomaly were rejected. After filtering, the total net
exposure time of each GIS was 16.2 ksec. 
We extracted the GIS spectra
from a circular region of radius $6'$ centered at the position of the
source, while the background was extracted from an annulus region
around the source. We have searched for any contamination from other
X-ray sources within $10'$; there are only a few very faint sources in
the {\it ROSAT} PSPC catalog (Supper et al. 1997) and in our {\it
Chandra} observation. Hence any contamination would be minimal.
Spectra were rebinned to have at least 20 photons in
each energy bin and we fitted the GIS2 and GIS3 spectra
simultaneously. We also set the normalization of GIS3 to be a free
parameter to account the difference in the calibration of the two
detectors. In all fits the GIS3 normalization relative to GIS2 is about 5\%.

\subsubsection{{\it Chandra} Observations}

\bol\ was located in the field of an ongoing (GO--AO2) X-ray census
of M31. It was observed on 2000
November 1 for 13.8 ksec. The Advanced CCD Imaging Spectrometer (ACIS-S)
was in the focal plane and \bol\ was located in the ACIS-S2 chip and was
$\sim 4.4'$ from the aim-point. Data were extracted from 0.3--7 keV to
minimize the background, and a circular extraction region of radius
$6''$ centered at the source position was used. Background was also
extracted from an annulus region centered on the source and it
contributed less than 0.1\%. The source spectrum was rebinned so that
there are at least 20 photons in each energy bin in order to allow
$\chi^2$ statistics. The response matrix
and the ancillary response file were generated by CIAO v2.1 \footnote{http://asc.harvard.edu/ciao/}. We note that
the \chandra\ spectrum suffers significant pile-up ($\sim 30\%$).

In addition, a sequence of short (0.8 ksec to 2 ksec [see Table 1])
High Resolution Camera (HRC)
exposures was available, because \bol\ happened to be imaged during
more than one year of monitoring carried out as part
of
a GTO program (see Garcia \etal\ 2000). 
In these images, \bol\ was $\sim 10'$ off-axis. 
We have extracted the light curve
of the source from a $25''$ circular region centered on the source position.

\subsection{Spectral Analysis and Results}

The spectra for the \rosat, \asca\ and \chandra\ data were fitted with
a variety of spectral models using XSPEC v11 \footnote{http://heasarc.gsfc.nasa.gov/docs/xanadu/xspec/index.html}. We first tried to fit
the spectra with single-component models
including absorbed power-law, black-body, thermal bremsstrahlung, disk
black-body and cut-off power-law models. Power-law models provide
 good fits to each data set, except the \chandra\ ACIS-S data. 
Derived values of photon index range from 1.3
to 1.7, with $N_H \sim (1-3)\times 10^{21}$ cm$^{-2}$. The
Galactic hydrogen column in the direction of M31 is about $7\times
10^{20}$ cm$^{-2}$ (Dickey \& Lockman 1990), and therefore our results
are consistent with additional local absorption, either due to M31
itself or to absorption within the system.
 An absorbed black-body model gave a value of $N_H$ ($\sim 3\times 10^{20}$
cm$^{-2}$) smaller than the Galactic value for all observations, and
we therefore discarded this model. We also fit the data to
thermal bremsstrahlung, cut-off
power-law, and disk black-body models; these fits
 gave very large uncertainties for the \rosat\ and \asca\
data, and the fit was unacceptable ($\chi^2_{\nu} \sim 1.4$) for the
\chandra\ data. 
Results of 
spectral fits to single-component models are shown in Table 2. The 
errors correspond to 90\% confidence levels for a single interesting parameter.

We next tried to fit the spectra with different two-component
models. A black-body plus power-law model gave an acceptable fit
($\chi^2_{\nu}= 1.27$) for the
\chandra\ data; for the \asca\ data, the additional black-body
component does not improve the fit significantly (at the 99\% level),
indicating that the black-body component is soft ($< 1$ keV). To
check the fit of the \chandra\ data, we also fitted the \rosat\ and
\asca\ data simultaneously so that we have broad energy coverage
from 0.1--10 keV. Since \bol\ is a variable, we choose the \rosat\
observation with flux level similar to that of the \asca\ data. The 0.5--2.4
keV fluxes of \rosat\#1 and \rosat\#2 were $\sim 4\%$ lower than the
\asca\ observation. \rosat\#2 is close to the `rib' and there
are shadowing effects.
We therefore used \rosat\#1 for the
joint spectral fittings. A single power-law model gave a reasonable
fit to the data, and a black-body plus power-law model did improve the
fit ($\chi^2_{\nu} = 1.16$). We replaced the
power-law component with a cut-off power law. However, it did not give a better
fit and the cut-off energy is larger than 10 keV, indicating
that the cut-off energy is much higher than 10 keV and is therefore  not
accessible to either \asca\ or \chandra. 

Note that the spectral fit to the {\it Chandra} data may be affected
by pile-up, which we have estimated to be at the 30\% level.
Normally one would expect pile-up to yield a somewhat harder spectrum
than the true emitted spectrum. In fact, the values of $\alpha$ and $k\, T$
derived form the {\it Chandra} data are in good agreement with
those derived from the combined {\it ROSAT}/{\it ASCA} fit, which does
not suffer from pile-up. The higher value of $N_H$
in the {\it Chandra} fit may, however, be related to the effects of pile-up.

We also employed a
disk black-body component instead of blackbody, it gave a
equally good fit ($\chi^2_{\nu} = 1.17$) but with larger errors on all
parameters. The derived inner radius of the accretion disk
$R_{in}\sqrt{\cos\theta}$ is $10\pm6.2$ km (assuming distance to M31 is
780 kpc; Stanek \& Garnavich 1998) with the disk temperature of
$1.95\pm0.35$ keV. In Table 2, we have summarized the best-fit
parameters.

\begin{figure*}
{\rotatebox{-90}{\psfig{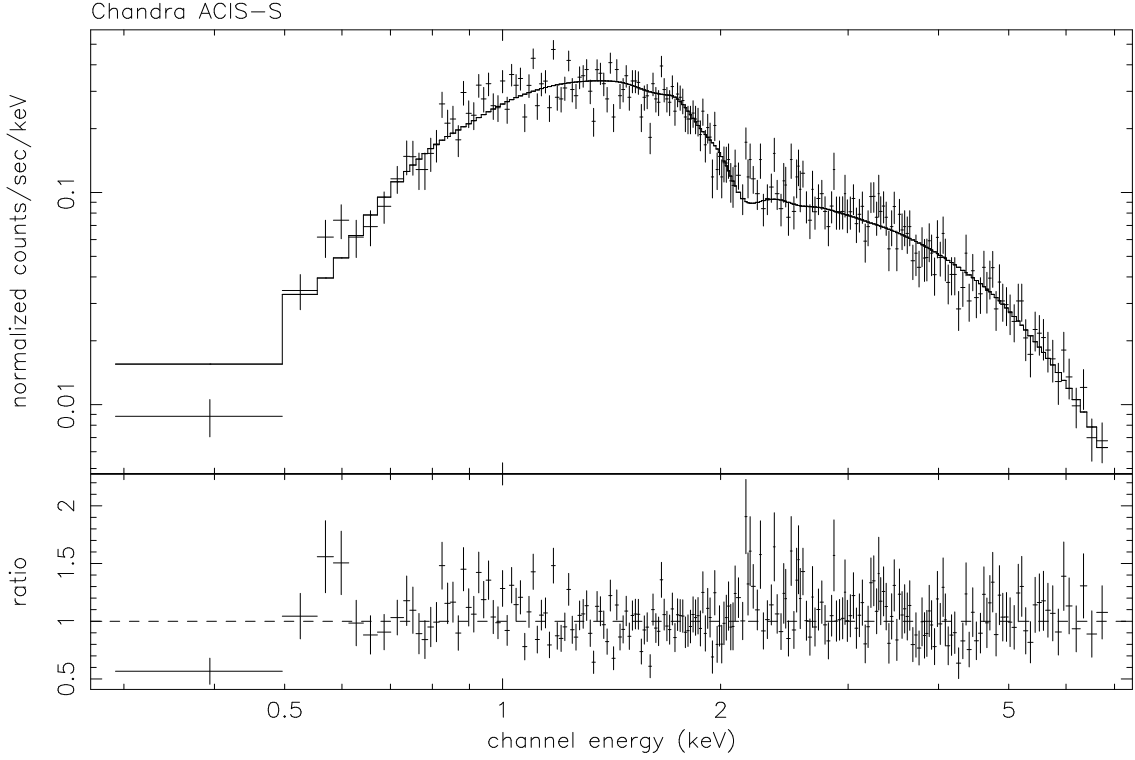}}}
\caption{{\it Chandra} ACIS-S spectral fit of \bol. 
The spectrum was
fitted by an absorbed power law plus blackbody model ($N_H=3.35\times
10^{21}$ cm$^{-2}$, $\alpha=1.67$ and $kT=0.80$ keV) The observed luminosity
(0.3--7 keV) at 780 kpc is $4.2\times10^{38}$ ergs s$^{-1}$.
}
\end{figure*}

Since the estimated metallicity of \bol\ is low (6\% solar; Huchra,
Brodie \& Kent 1991), we employed a photoelectric model with variable
abundances (VPHABS in XSPEC), together with a photoelectric absorption
fixed at the value towards M31, $7\times10^{20}$ cm$^{-2}$. We grouped
H and He as one parameter fixed at the cosmic value, and set other metals
(C, N, O, Ne, Na, Mg, Al, Si, S, Cl, Ar, Ca, Cr, Fe, Ni, and Co) as
another parameter, fixing the abundance at 0.06 of the solar
value. The black-body plus power-law model of the \chandra\ data was improved
($\chi^2_{\nu}=1.29/206\,dof$; $N_{H_{vphabs}}=\left(4.33\pm0.67\right)\times10^{21}$
cm$^{-2}$; $\alpha=1.39\pm0.10$; $kT=0.76\pm0.05$ keV). However, there is no
improvement on the joint \rosat\ and \asca\ data and the parameters
are roughly the same as before.

\begin{figure*}
{\rotatebox{-90}{\psfig{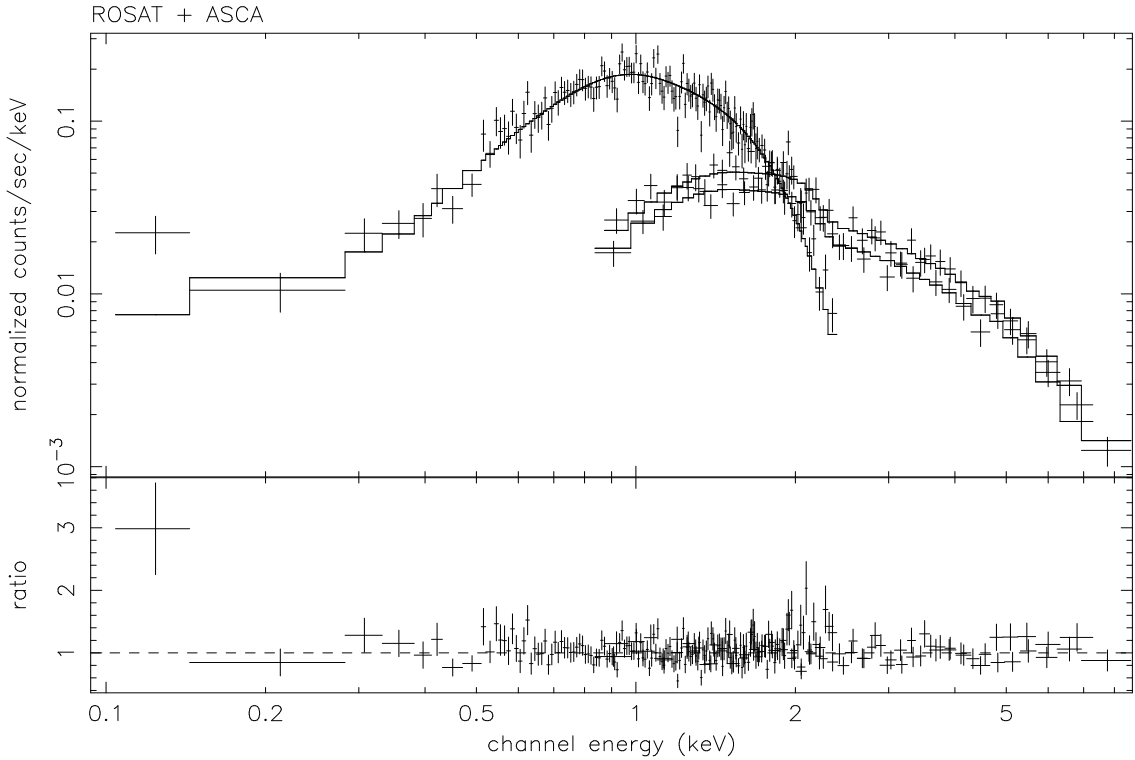}}}
\caption{Joint \rosat\ and \asca\ spectral fit of \bol. 
The spectrum
was fitted by an absorbed power law plus blackbody model ($N_H=1.61\times
10^{21}$ cm$^{-2}$, $\alpha=1.61$ and $kT=0.86$ keV) The observed luminosity
(0.3--7 keV) at 780 kpc is $6.1\times10^{38}$ ergs s$^{-1}$.
}
\end{figure*}

{\small
\begin{table*}
\begin{center}
TABLE~2

{\sc Best-fit parameters for the energy spectra of \bol}

\label{tab:spectra}
\vspace{1mm}
\begin{tabular}{c c c c c c c}
\tableline \tableline
\noalign{\smallskip}
Observation & $N_H$ & $\alpha$ & $kT$~\tablenotemark{a} & $R_{in}\sqrt{\cos\theta}$~\tablenotemark{b} &
$\chi^2_{\nu}/dof$ & Luminosity~\tablenotemark{c} \\
& ($10^{21}$ cm$^{-2}$) & & (keV) & (km) & & \\
\hline\\
\rosat\#1 & $0.96\pm0.96$ & $1.33\pm 0.08$ & && 1.32/156 & 2.16 \\
\rosat\#2 & $0.98\pm0.15$ & $1.30\pm 0.11$ & && 1.01/147 & 2.34 \\
\rosat\#3 & $1.10\pm0.20$ & $1.32\pm 0.12$ & && 0.90/148 & 1.43 \\
\asca\   & $2.20\pm0.60$ & $1.70\pm 0.07$ & && 0.90/81  & 2.25 \\
\chandra\ & $3.91\pm0.16$ & $1.67\pm0.03$& && 1.53/209 & 1.42  \\
& $3.35\pm0.34$ & $1.67\pm0.13$ & $0.80\pm0.11$&& 1.40/207 & 1.44\\
\rosat\#1+\asca & $1.20\pm0.90$ & $1.58\pm0.04$ & && 1.22/238 & 2.29\\
& $1.61\pm0.08$ & $1.61\pm0.08$& $0.86\pm0.12$ & & 1.16/236 & 2.29\\
                  & $1.10\pm0.20$ & $1.77\pm 0.34$ & $1.95\pm0.35$ & $10\pm6.2$
         & 1.17/236 & 2.28 \\
\noalign{\smallskip}
\tableline
\\
\multicolumn{7}{l}{NOTES --- The power-law model is defined as $AE^{-\alpha}$,
where $A$ is the normalization at 1 keV}\\
\multicolumn{7}{l}{(photons cm$^{-2}$ s$^{-1}$ keV$^{-1}$) and $E$ is in keV.}\\
\multicolumn{7}{l}{$^a$ Black-body temperature or inner disk temperature}\\
\multicolumn{7}{l}{$^b$ Assuming a distance of 780 kpc}\\
\multicolumn{7}{l}{$^c$ Emitted 0.5--2.4 keV luminosity in units of $10^{38}$
erg s$^{-1}$; we choose this energy range for convenient}\\
\multicolumn{7}{l}{comparison with \rosat\ data.}
\end{tabular}
\end{center}
\end{table*}
}

We have performed spectral analysis of \bol\ using extensive \rosat\,
\asca\ and \chandra\ data. The spectral characteristics resemble those
seen in low-mass X-ray binaries (LMXBs). The source spectra can be well fitted by a
single power-law. However, for data which span the widest
range of energies, combined \rosat\ and
\asca\ data and recent \chandra\ observations, an additional black-body
component improves the fit. Such a two-component model was previously
used to study the spectrum of many LMXBs including GC sources
X\,2127+119, X\,1746--370 and X\,1724--308 (e.g. Christian \& Swank
1997; Guainazzi \etal\ 1998; Church \& Balucinska-Church 2001)
in our Galaxy. Our derived photon index and black-body temperature for
\bol\ are about 1.6 and 0.8 respectively, which is similar to that of
X\,2127+119 (Christian \etal 1997) and X\,1724--308 (Guainazzi \etal 1998).

There are several possible physical explanations for the spectral model. 
The soft  thermal component (i.e. black-body component)
could originate from either an optically thick accretion disk or an
optically thick boundary layer where the accreted matter interacts with
the neutron star surface (see, e.g. White, Stella \& Parmar 1988).
The power-law component can be formed by scattering 
of the soft
black-body photons in an extended region above the accretion
disk, 
or from an optically thin boundary layer (Guainazzi
\etal\ 1998; Barret \etal\ 2000).

\subsection{Why is Bo\,375 so luminous?}

The spectral fits tell us that the 
luminosity of \bol\ has ranged from $\sim 2-6\times10^{38}$ ergs
s$^{-1}$ during the past 20 years.
This makes it more luminous than any Galactic GC source by a factor
that can be larger than $10.$ 
The source is also
more luminous than the other GC sources we have so far
studied  in M31, but by a much smaller factor (approximately
a factor of $2$).  

\subsubsection{Is Bo\,375 an unusual globular cluster?}

Bo\,375 (also known as G307) is not an unusual M31 \gc .
The values of its measured parameters are very close to the  median values
for M31's \gc\ system: $V= 17.62$  ($17.04$ is the median value);
$B-V=0.90$ ($0.90$ is the median value);  
$V-R=0.54$ ($0.56$ is the median value);  
$V-I=1.02$ ($1.20$ is the median value). 
Its radial velocity is $-196 +/- 70$ km/s, while $-280$ km/s is the 
median value.
[Fe/H]= $-1.23 +/- 0.22$, with  $-1.21$ the median value. 
Thus, aside from being a bit fainter and having a somewhat less negative
radial velocity, Bo\,375 is incredibly typical.
In a coordinate system centered on M31, with $X$ equal to the
distance  along major axis and $Y$ equal to the
distance  along minor axis, it's at $X=40', Y=-12'$; i.e.,
it is projected just inside the disk boundaries.

It is also important to compare the properties of Bo375 to those of the
Galactic globular clusters that house X-ray sources. Combining the lists
of Verbunt et a.\ (1995) and Sidoli et al (2001), we compiled a list of
14 Milky Way GCs with bright X-ray sources and extracted their properties from
the 1999 edition of the Harris (1996) catalog. Bol 375 is more metal-poor than
most Milky Way GCs with X-ray sources (median ${\rm [Fe/H]}=-0.50$,
and less optically luminous (Bol375 $M_V=-7.2$; median MW XRS $M_V=-7.7$.
It is also further from the center of M31 than most Milky Way X-ray source clusters
are from the center of the Galaxy. However, all of the properties of Bol375
fall within the range of properties seen in Milky Way X-ray source clusters,
so there is no obvious way to explain what causes its high X-ray luminosity.
Measurements of the structural properties of Bol375, which might provide
and important clue, are unfortunately unavailable.

\subsubsection{Is it a Multiple Source?} 

We will argue in \S 6 that multiple X-ray sources are expected among
the M31 globular clusters. Thus, the most natural explanation for a
high-luminosity source in an M31 \gc\ may be that the source is
actually a composite of $2$  or more sources.
We have therefore studied the image (composed of hits by $\sim 7000$
photons) of Bo\,375 to see if it
suggests multiplicity. Indeed, the image breaks into
$2$ components, but we find that this is most likely due to the
shape of the point spread function (PSF) 
at the position of the source. Nevertheless, 
there appears to be some residual broadening of the image along the direction
perpendicular to the PSF-induced binarity. 
Further checks of the PSF, especially attuned to the spectrum of Bo\,375,
are necessary to establish that the broadening is a real effect.
While such work is underway, we note that the maximum separation between
components consistent with our observations is 
$\sim 1-2''$, which corresponds to a physical projected separation 
of $\sim 3.7-7.4$ pc. Such a separation could be consistent
with a model in which there are two or three X-ray bright binaries
in the cluster, and one or two of these binaries was ejected from the 
cluster core $\sim 10^5$ years ago.
Such separations, and even smaller ones,
could be
resolved by an on-axis observation with the HRC. Depending on the
nature of the sources, HST measurements
could also prove to be valuable. 

The ejection of X-ray binaries from the core of a globular cluster,
achieving $\sim pc$ separations while the X-ray evolution continues,
is consistent with physical models of X-ray
sources in \gc s.
The basic picture that has emerged
to explain the overabundance (by a factor of $\sim 100$; Clark 1975) of
X-ray sources in \gc s, is the following. Stars with the 
highest masses, and also primordial binaries,
migrate toward the cluster centers. Stellar densities near the center
are typically so high ($10^3-10^6$ stars pc$^{-3}$) that individual stars
and binaries can interact
via two-body, three-body ($2\ +\ 1$) and four-body ($2\ +\ 2$)
interactions. Such interactions can produce close binary systems
which subsequently experience mass transfer and are luminous at X-ray
wavelengths. These close binaries store a great deal of potential energy;
if they interact with other stars or binary systems, they (or the systems
with which they have interacted) can be ejected from the core and
even from the \gc . Thus the environment that favors the formation
of X-ray binaries in or near the cores of \gc s also  leads them
to eventually be ejected from the cores or even from the
clusters.

For Bo\,375, however, multiplicity is likely to be, at  most,
only part of the story. As described in \S 2.4.3, there is significant
time variability over timescales ranging from 16 hrs to many months;
this indicates that one source must contribute a large fraction
(although possibly not all) of the detected flux.

\subsubsection{Is it a Black Hole?} 
When observing a source which is more luminous than the Eddington
limit for a $1.4 M_\odot$ object, it is natural to conjecture that
the accreting source may be a black hole.
Reliable detections of solar-mass black holes have so far
been based on orbital studies that find upper limits on the mass
of an accreting compact object well above the maximum mass
predicted for neutron stars (see, e.g., Charles 1998). Such a study is not
presently feasible for Bo\,375. We must therefore search
for spectral signatures that could possibly distinguish
between neutron star and black hole models. 
First-principles calculations (e.g. from ADAF models; see Narayan,
Barret, \& McClintock 1997)
and comparisons between the spectra of neutron star accretors and
dynamically-confirmed black hole accretors can provide
some insight. No reliable set of tests is
presently available, however, to allow us to place Bo\,375 firmly in either the
``black hole" or ``neutron star" camp.  

The discussion at the end on \S 2.3 indicates that there are similarities
between the spectra of MW GC X-ray sources, which are known to be accreting
neutron stars, and the spectrum of Bo\,375. Such broad similarities
are, however, only suggestive and cannot definitively establish the
nature of the source.

The time variability (discussed below) may also indicate that nature
of the accreting object.  While Bol\,375 is highly variable, it is not
a transient source.  Within the MW, black hole binaries can be
separated into those that are transient, and those that are
persistent.  The persistent sources have high mass secondaries (generally
above $5 M_\odot$), while
the transient systems tend to have low mass secondaries (see Tanaka
\& Shibazaki 1996, Table 2).  Given the location of this source in a
GC, the secondary is likely to be of low mass.  Its non-transient nature
and likely low-mass secondary could then be taken to
argue against the black hole hypothesis.
However, we add the following caveat to this argument.
The transient nature of the systems with low mass secondaries is
likely caused by mass transfer rates which are below a critical value for
stability (van Paradijs 1996).  Even low mass secondaries can briefly
supply high mass transfer rates when they begin moving off the main
sequence (see section 6.1 herein). Should Bol 375 be in this unusual
state, its persistent nature would not indicate the nature of the
primary.

\subsubsection{Time Variability}

The neutron-star nature of the MW GC X-ray sources was definitively established
by observations of time variability. Specifically, X-ray bursts have been
observed from all globular cluster X-ray sources. These bursts 
can be linked to the episodic nuclear burning of matter accreted by the
neutron star.    
An X-ray source can also be established as a neutron star system if it
is observed to emit periodic X-ray pulses. 

Given the important role time variability studies have played 
in helping us to understand Galactic GC X-ray sources, it is important to  
study the time variability of Bo\,375. We have used the GO data set to
study its short-term behavior. (Note that the count rate ($\sim 0.5$ counts/s)
is too low to allow us to probe time scales much shorter than a minute.) 
Fourier analysis does not find evidence of periodicity, and there is no
evidence of bursts. Thus, the light curve does not allow us to establish
that all or part of the X-radiation  is emitted by a neutron star system.

In Figure 4 we show the light curves of data binned into intervals of
$0.5$ min, $1$ min, $5$ min, and $10$ min. 
Data from Bo\,375 is displayed on the left. Let $A_{ti}$ be the average
number of counts in the time interval under consideration. The upper (lower)
dashed line is positioned at $A_{ti}+ \sqrt{A_{ti}}$ ($A_{ti}- \sqrt{A_{ti}}$)
These two dashed lines thus provide guidelines to the uncertainty due
to the counting statistics. The dark uneven curve that generally
lies between these two dashed lines is the running average, 
computed by considering
the central bin and the four previous and subsequent bins.
At short time scales, there is no evidence for time variability.

\begin{figure*}[t]
\begin{center}
\psfig{file=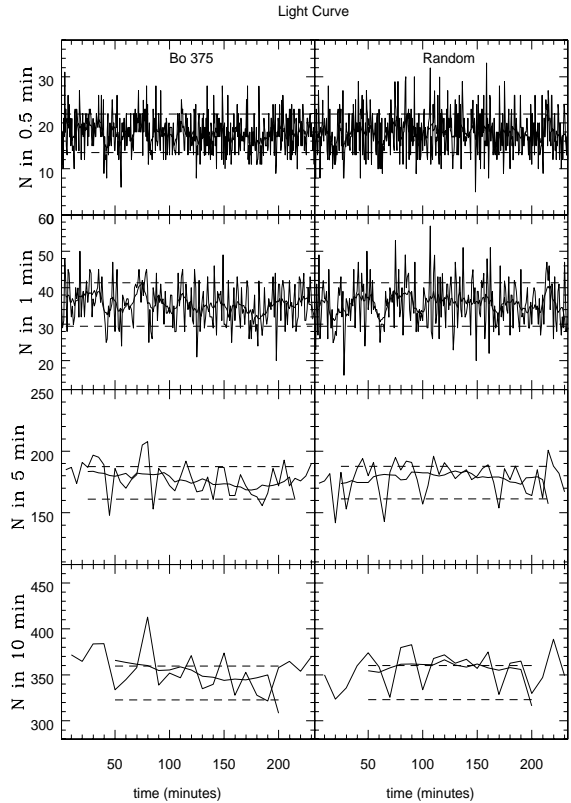,height=11cm,width=8cm}
\end{center}
\caption{The light curves of Bo\,375 binned into intervals of 0.5 min, 1
min, 5 min and 10 min. Clearly there is no evidence for time
variability on the time scale of the {\it Chandra} observation. See \S 2.4.4 for discussion.}
\end{figure*}

The light curves plotted on the right are also taken from the Bo\,375 data, but the 
order of the data points (taken in $5$ s time bins) has been randomized.
There are no readily quantifiable differences between the randomized and
correctly-time-ordered  data.

\begin{figure*}[t]
\begin{center}
{\rotatebox{-90}{\psfig{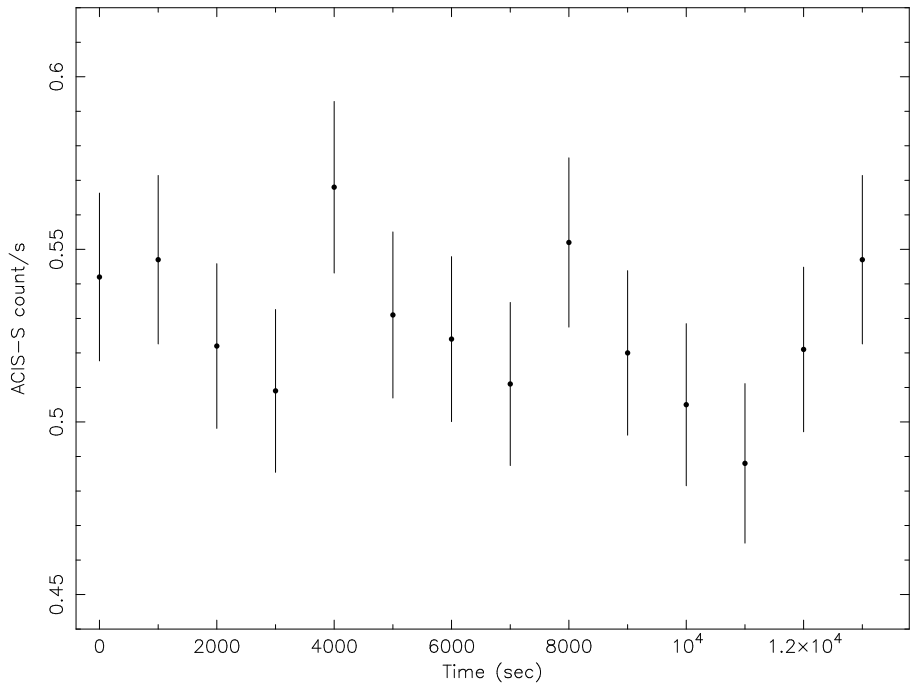}}}
\end{center}
\caption{15 ksec \chandra\ ACIS-S light curve of \bol. The time resolution is
1000 sec.}
\end{figure*}

Whereas these observations covered
a time interval of roughly $4$ hours, the 1991 {\it ROSAT} survey of M31
has data on Bo\,375 that spans a baseline that is  roughly $10$ times
longer. Supper et al. (1997) find evidence for regular variations
of $\sim 50\%$ on a time scale of $\sim 16$ hours. Proceeding to
longer time scales, Bo\,375 was in the fields visitied
at regular ($\sim$ monthly) intervals by {\it Chandra's}
HRC for $\sim 1$ ksec observations. This data set also
finds significant variations from $50-100\%$ (Figure 6).

\begin{figure*}[t]
\begin{center}
\psfig{file=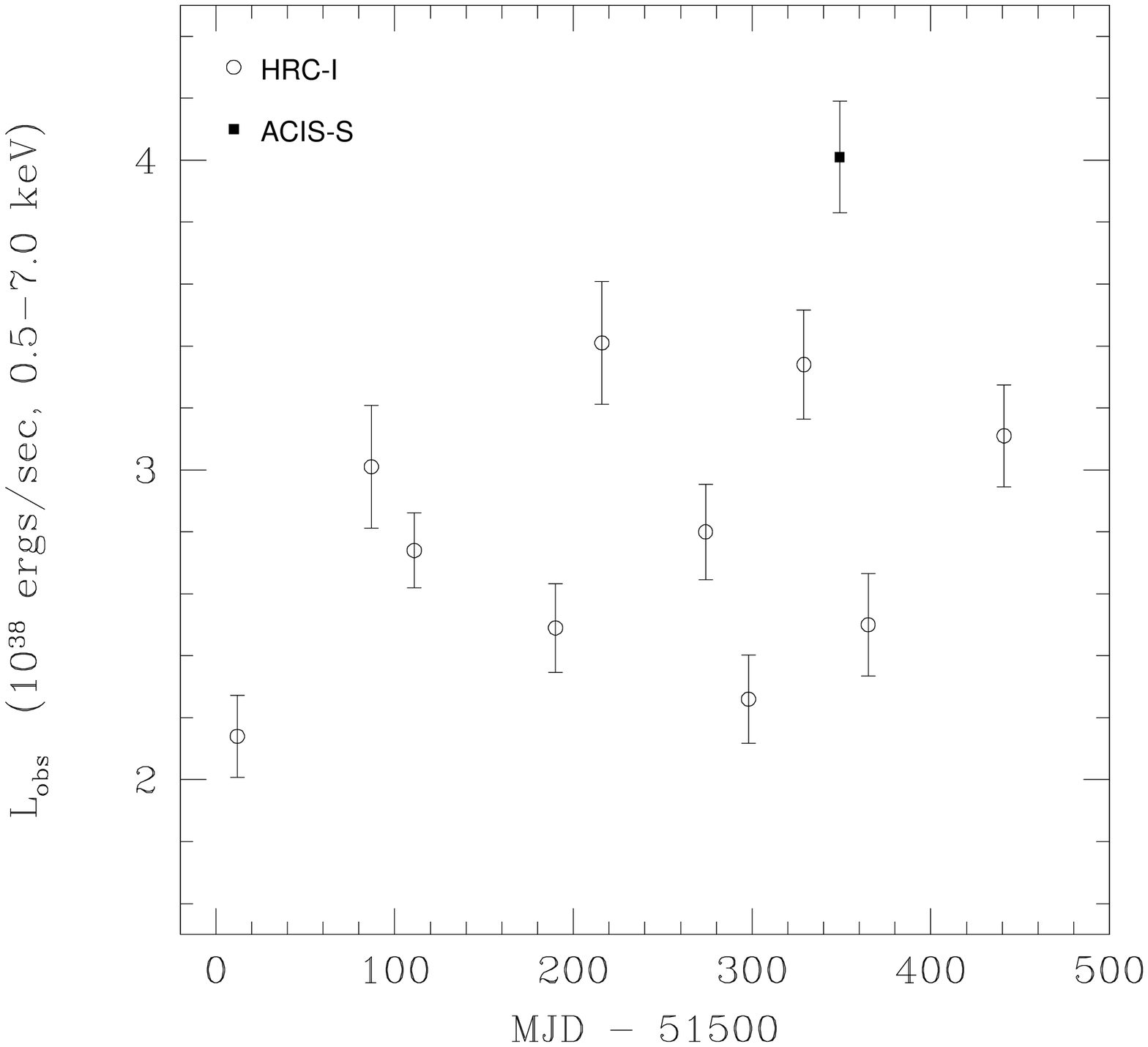,height=10cm,width=10cm}
\end{center}
\caption{\chandra\ HRC light curve of \bol. The luminosity was
estimated by PIMMS, using the observed count rate and assuming a power law spectral model with $N_H=10^{21}$
cm$^{-2}$ and photon index $\alpha=1.7$.}
\end{figure*}

In summary, what we know so far about Bo\,375's time variability provides
no hint that the X-ray source is a neutron star. The data
do suggest, however, that if the Bo\,375 source is
a multiple, the number of separate bright components is small.
This implies that there is at least one X-ray binary whose
luminosity appears to be comparable to the Eddington limit for a $1.4\, M_\odot$
accretor. We find that a model in which the luminosity is generated by
mass transfer onto a neutron star, occurring on the thermal time
scale of the donor, is consistent
with this data (\S 6). Alternatively, 
the high luminosity could be an effect of beaming, or it
could be associated with a high-mass, possibly a black hole, accretor.

\section{Other Bright M31 GC Sources}

\subsection{Brief Description of Data} 

Our \chandra\ census fields (see Figure 1) contained other bright X-ray
sources also in GCs. This suggested that it would be valuable to
include additional M31 fields which had already been observed by
\chandra. Such data was collected by a GTO program (see Garcia
et al. 2000; Primini et al. 2000). There were seven ACIS-I pointings
of the central $16'\times16'$ region around the galaxy from 1999-2000. 
For the analysis below, we used 8.8 ksec from the longest pointing
(1999 October 13), except that two additional GCs were observed on
1999 December 11 and 2000 January 29.

\subsection{The Sources}

Table 3 provides a complete list of all of the X-ray sources we have observed
in M31 globular clusters. The sources are listed in order of
count rate; the source with the highest count rate appears first.
For each source, the coordinates (J2000), optical ID and luminosity
(assuming a power law model with $N_H=10^{21}$ cm$^{-2}$ and photon
index $\alpha=1.7$) are included. Three sources (Source 2, 13 and 15)
have been observed by two GO pointings and we list both observations
with the first one corresponding to the earlier observations. Table 4 lists the counts in a sequence of energy bins.

\begin{table*}
\footnotesize
\begin{center}
TABLE~3

{\sc Globular Cluster X-ray Sources in M31}

\vspace{1mm}
\begin{tabular}{lcccccl}
\tableline \tableline
\noalign{\smallskip}
ID&Source Name& R.A. & Dec. & Exposure& $L_X$(0.3--7 keV)& Optical ID\\
& & (h:m:s) & ($^\circ:':''$) & (ksec) &($10^{37}$erg s$^{-1}$)&\\
\hline
1$^\dagger$&CXOM31 J004545.4+413942 & 00:45:45.4& +41:39:42 &13.8&
29.94&Bo\,375, mita468\\
2$^{\dagger \diamondsuit}$&CXOM31 J004215.8+410114 & 00:42:15.8&
+41:01:14 &14.5& 11.97&Bo\,82, mita159\\
 &                      &                         &   & 13.0   & 8.90   & \\
3$^\dagger$&CXOM31 J004310.6+411450 & 00:43:10.6& +41:14:50 &8.8&
8.43&Bo\,153, mita251\\
4$^\dagger$&CXOM31 J004218.5+411400 & 00:42:18.5& +41:14:00 &8.8&
4.69&Bo\,86, mita164\\
5$^\dagger$&CXOM31 J004259.6+411918 & 00:42:59.6& +41:19:18 &8.8& 2.79&Bo\,143\\
6&CXOM31 J004259.8+411604 & 00:42:59.8& +41:16:04 &8.8& 2.68&Bo\,144\\
7$^\dagger$&CXOM31 J004302.9+411521 & 00:43:02.9& +41:15:21 &8.8& 2.18&Bo\,146\\
8$^\dagger$&CXOM31 J004303.8+411804 & 00:43:03.8& +41:18:04 &8.8& 1.83&Bo\,148\\
9$^\dagger$&CXOM31 J004337.3+411442 & 00:43:37.3& +41:14:42&4.1& 1.83&mita299\\
10&CXOM31 J004209.4+411744 & 00:42:09.4& +41:17:44 &8.8& 1.17&mita140\\
11$^\dagger$&CXOM31 J004303.2+412121 & 00:43:03.2& +41:21:21 &8.8&
1.09&Bo\,147, mita240\\
12$^\dagger$&CXOM31 J004314.4+410726 & 00:43:14.4& +41:07:26 &4.1& 0.91&Bo\,158\\
13$^{\dagger \diamondsuit}$&CXOM31 J004206.1+410248 & 00:42:06.1&
+41:02:48 &14.5& 0.85& Bo\,D42, mita130\\
  &                      &                        &    & 13.0   &  0.74   & \\
14&CXOM31 J004226.0+411914 & 00:42:26.0& +41:19:14 &8.8&
0.74&Bo\,96, mita174\\
15$^{\diamondsuit}$&CXOM31 J004207.1+410016 & 00:42:07.1& +41:00:16 &14.5& 0.71& Bo\,D44\\
  &                      &                        &    & 13.0   & 0.31    & \\
16$^\dagger$&CXOM31 J004231.2+411938 & 00:42:31.2& +41:19:38 &8.8&
0.44&Bo\,107, mita192\\
17&CXOM31 J004246.0+411735 & 00:42:46.0& +41:17:35 &8.8& 0.31&PB-in7$^{\clubsuit}$\\
18&CXOM31 J004240.6+411031 & 00:42:40.6& +41:10:31 &8.8&
0.28&Bo\,123, mita212\\
19&CXOM31 J004307.5+412019 & 00:43:07.5& +41:20:19 &8.8&
0.27&Bo\,150, mita246\\
20&CXOM31 J004212.0+411757 & 00:42:12.0& +41:17:57 &8.8&
0.26&Bo\,78, mita153\\
21$^\dagger$&CXOM31 J004225.0+405719 & 00:42:25.0& +40:57:19 &13.0&
0.17& Bo\,94, mita173\\
22&CXOM31 J004152.9+404710 & 00:41:52.9& +40:47:10
&13.0&0.13&Bo\,58, mita106\\
23&CXOM31 J004241.4+411522 & 00:42:41.4& +41:15:22 &8.8& 0.11 &mita213\\
24*&CXOM31 J004219.6+412153 & 00:42:19.6& +41:21:53 &8.8&
0.08&mita165, mita166\\
25&CXOM31 J004227.4+405936 & 00:42:27.4& +40:59:36 &13.0& 0.08& Bo\,98\\
26&CXOM31 J004315.4+411124 & 00:43:15.4& +41:11:24 &8.8&
0.06&Bo\,161, mita260\\
27&CXOM31 J004141.0+410401 & 00:41:41.0& +41:04:01 &13.0& 0.06& mita87\\
28&CXOM31 J004250.7+411032 & 00:42:50.7& +41:10:32 &8.8&
0.05&mita222, PB-in2$^{\clubsuit}$\\
\tableline
\noalign{\smallskip}\\
\multicolumn{7}{l}{NOTES --- Observations lasting longer than 10 ksec
were carried out as part of the GO program (see Figure 1); Shorter}\\
\multicolumn{7}{l}{observations around the central
$16'\times16'$ field were carried out as part of the GTO program (Garcia
et al. 2000;}\\
\multicolumn{7}{l}{Primini et al. 2000).}\\
\multicolumn{7}{l}{$\dagger$ Source detected by \rosat\ (Supper et al. 1997)}\\
\multicolumn{7}{l}{$\diamondsuit$ Source detected in both GO observations}\\
\multicolumn{7}{l}{$\clubsuit$ Results based on an HST survey (Barmby
\& Huchra 2001)}\\
\multicolumn{7}{l}{* Source 25 is associated with two clusters: mita165
and mita166}
\end{tabular}
\end{center}
\end{table*}

\begin{table*}
\footnotesize
\begin{center}

TABLE~4

{\sc Counts in different energy bands of Globular Cluster X-ray Sources in M31}

\vspace{1mm}
\begin{tabular}{lccccccc}
\tableline \tableline 
\noalign{\smallskip}
Source&  & &Counts & & & &Hardness\\
 & 0.1--0.4 keV&0.5--0.9 keV&0.9--1.5 keV&1.5--2 keV&2--4 keV&4--10
 keV &Ratio$^a$\\
\hline
1&$0.0\pm0.0$&$593.8\pm24.4$&$2319.1\pm48.2$&$1509.0\pm38.8$&$2060.8\pm45.4$&$897.4\pm30.1$&1.0\\
2&$9.3\pm3.3$&$178.8\pm13.4$&$817.7\pm28.6$&$633.3\pm25.1$&$993.6\pm31.5$&$423.0\pm20.7$&1.4\\
 &$7.4\pm3.1$&$135.4\pm11.7$&$509.4\pm22.7$&$461.9\pm21.5$&$740.9\pm27.3$&$425.4\pm20.8$&1.8\\
3& $0.7\pm1.0$&$121.8\pm11.0$&$322.5\pm18.0$&$163.3\pm12.7$&$233.5\pm15.3$&$123.3\pm11.1$&0.8\\
4& $1.0\pm1.0$&$66.3\pm8.1$&$186.0\pm13.6$&$104.3\pm10.2$&$128.3\pm11.3$&$57.4\pm 7.6$&0.7\\
5&$1.0\pm1.0$&$54.4\pm7.4$&$126.1\pm11.2$&$52.9\pm7.2$&$65.1\pm8.1$&$21.7\pm4.7$&0.5\\
6& $1.0\pm1.0$&$35.8\pm6.0$&$91.0\pm9.5$&$58.6\pm7.6$&$75.9\pm8.7$&$44.5\pm6.7$&0.9\\
7&$1.0\pm1.0$&$36.7\pm6.0$&$88.0\pm9.3$&$48.4\pm6.9$&$59.1\pm7.7$&$20.8\pm4.5$&0.6\\
8&$1.0\pm1.0$&$29.7\pm5.4$&$87.7\pm9.3$&$43.8\pm6.6$&$32.0\pm5.6$&$14.7\pm3.8$&0.4\\
9&$1.0\pm1.0$&$8.2\pm3.3$&$33.8\pm5.9$&$11.3\pm3.7$&$27.3\pm5.4$&$14.1\pm4.3$&1.0\\
10&$0.0\pm0.0$&$8.0\pm2.8$&$56.9\pm7.6$&$30.5\pm5.5$&$28.1\pm5.3$&$12.3\pm3.5$&0.6\\
11& $0.0\pm0.0$&$24.0\pm4.8$&$56.3\pm7.5$&$16.3\pm4.0$&$20.2\pm4.5$&$9.3\pm3.1$&0.3\\
12&$0.0\pm0.0$&$1.0\pm1.0$&$8.0\pm2.8$&$6\pm2.4$&$22.0\pm4.8$&$10.5\pm3.3$&3.6\\
13&$11.0\pm3.4$&$74.3\pm8.6$&$83.3\pm9.1$&$26.6\pm5.1$&$30.2\pm5.5$&$5.0\pm3.0$&0.2\\
  &$5.4\pm2.4$&$52.7\pm7.2$&$65.4\pm8.1$&$23.0\pm4.7$&$25.7\pm5.1$&$8.2\pm3.0$&0.3\\
14&$0.0\pm0.0$&$11.0\pm3.3$&$37.5\pm6.1$&$15.8\pm3.9$&$14.9\pm3.8$&$7.0\pm2.6$&0.4\\
15& $1.0\pm1.4$&$24.3\pm5.0$&$65.3\pm8.1$&$35.6\pm6.0$&$40.2\pm6.4$&$20.0\pm4.9$&0.6\\
  &$0.0\pm0.0$&$16.7\pm4.1$&$21.0\pm4.5$&$13.0\pm3.6$&$18.0\pm4.2$&$9.2\pm3.1$&0.7\\
16& $0.0\pm0.0$&$10.3\pm3.2$&$15.7\pm4.0$&$10.4\pm3.2$&$12.3\pm3.6$&$2.7\pm1.7$&0.6\\
17&$0.0\pm0.0$&$1.8\pm1.4$&$14.5\pm3.8$&$3.8\pm2.0$&$13.0\pm3.6$&$2.6\pm1.7$&0.9\\
18&$0.0\pm0.0$&$4.7\pm2.2$&$14.0\pm3.7$&$3.7\pm2.0$&$4.9\pm2.3$&$6.1\pm2.5$&0.6\\
19&$0.0\pm0.0$&$6.0\pm2.4$&$13.5\pm3.8$&$6.0\pm2.4$&$5.1\pm2.3$&$2.0\pm1.4$&0.4\\
20&$0.0\pm0.0$&$2.0\pm1.4$&$5.5\pm2.4$&$8.0\pm2.8$&$9.0\pm3.2$&$3.0\pm1.7$&1.6\\
21&$0.0\pm0.0$&$6.0\pm2.4$&$16.0\pm4.0$&$7.0\pm2.6$&$8.0\pm2.8$&$5.2\pm2.4$&0.6\\
22&$0.0\pm0.0$&$0.5\pm1.0$&$10.8\pm3.7$&$5.1\pm2.5$&$5.8\pm2.9$&$10.1\pm3.3$&1.4\\
23&$0.0\pm0.0$&$0.0\pm0.0$& $7.1\pm2.8$& $2.0\pm1.4$& $3\pm1.7$& $1.4\pm1.2$&0.6\\
24&$0.0\pm0.0$&$0.0\pm0.0$&$1.0\pm1.9$&$0.0\pm0.0$&$6.0\pm2.4$&$3.0\pm1.7$&9.0\\
25&$0.0\pm0.0$&$2.7\pm1.7$&$10.0\pm3.1$&$4.0\pm2.0$&$3.0\pm1.7$&$0.4\pm1.0$&0.3\\
26&$0.0\pm0.0$&$1.5\pm1.4$&$3.0\pm2.1$&$3.0\pm1.7$&$0.0\pm0.0$&$0.1\pm1.2$&0.02\\
27&$0.0\pm0.0$&$0.5\pm1.0$&$9.5\pm3.1$&$3.1\pm2.0$&$1.5\pm1.4$&$0.5\pm2.6$&0.2\\
28&$0.0\pm0.0$&$2.8\pm1.7$&$0.9\pm1$&$0.0\pm0.0$&$0.7\pm1.0$&$2.0\pm1.4$&0.7\\

\noalign{\smallskip}
\tableline
\\
\multicolumn{8}{l}{$^a$ Hardness ratio is defined as 2--10 keV/0.5--1.5 keV}
\end{tabular}
\end{center}
\end{table*}

\noindent{\it Optical IDs:} A wavelet detection algotithm ({\sc
WAVDETECT} in CIAO) was used
to find X-ray sources. Each {\it Chandra} source was placed 
at the center of a $3''$ source region. We cross-correlated with 
catalogs of globular clusters (Battistini et al. 1987; Magnier 1993;
Barmby et al. 2000; Barmby \& Huchra 2001).
Typical offsets between the globular clusters
and our X-ray sources are about $1''$; when a cluster appears
in multiple catalogs, the offsets may differ somewhat.

\subsection{Hardness Ratios and Spectra}

In order to quantify the spectral behavior of M31 X-ray GCs, we
complied a table for the counts in different energy bins (see Table 4).
The bins we use are $0.1-0.4$, $0.5-0.9$, 
$0.9-1.5$, $1.5-2.0$, $2-4$, and $4-10$ keV, respectively. The choice of
ranges for the low-energy bins is based on the {\it ROSAT} bins typically
used for these data sets (Verbunt \etal\ 1995, Supper \etal\ 1997), 
so as to facilitate comparisons between {\it Chandra}
and {\it ROSAT} data sets. The count rates in each bin are background subtracted.      

We implemented a set of spectral fits for each source with more than $\sim 400$
counts (0.3--7 keV).  There are $4$ such sources (in addition to Bo\,375) and a total of $5$ observations
with such count rates. They are: Bo 82 with 3053 counts in the first GO pointing and
2277 counts in the second pointing, Bo 153 with 962 counts in a GTO pointing with
ACIS-I, Bo 86 with 542 counts (also in a GTO pointing with ACIS-I),
and Bo 143 with 446 counts.

We first tried to fit the spectra with a power law model with the
hydrogen column density $N_H$ as a free parameter. If the derived
$N_H$ was inconsistent with the observed color excess $E(B-V)$ from
optical observations, then we fixed the $N_H$ according to
the optical absorption. We also tried to fit the spectra with other single
component models such as blackbody and thermal bremsstrahlung model. 
All of the fits were either unacceptable ($\chi^2_{\nu} >2$), or also the best-fit
parameters were unrealistic. Finally, we fitted
the spectra with power law plus blackbody models. Although these generally
gave good fits, the blackbody temperature became unrealistic and the
fits were not improved significantly. We describe below the best fit
parameters of power law models.

\noindent{\bf Bo 82} -- A power law model provied a good fit ($\chi^2_{\nu}$=1.03/114
dof) for the first GO observation with $N_H=(5.17\pm0.31)\times
10^{21}$ cm$^{-1}$ and $\alpha=1.42\pm0.06$. The emitting luminosity
in 0.3--7 keV is $1.70\times10^{38}$ erg s$^{-1}$. For the second GO
observation, we obtained $N_H=(4.58\pm0.40)\times
10^{21}$ cm$^{-1}$ and $\alpha=1.14\pm0.07$, with
$\chi^2_{\nu}$=1.16/88 dof. The luminosity is $2.20\times10^{38}$ erg
s$^{-1}$. 

\noindent{\bf Bo 153} -- The best fit parameters were $N_H=(1.07\pm0.33)\times 10^{21}$ cm$^{-1}$ and
$\alpha=1.38\pm0.10$, with $\chi^2_{\nu}$=0.92/40 dof. The luminosity
in 0.3--7 keV is $9.12\times10^{37}$ erg s$^{-1}$. Fixing the
hydrogen column density $N_H=6\times 10^{20}$ cm$^{-2}$
[$E(B-V)=0.11$] gave
$\alpha=1.26\pm0.05$ with $\chi^2_{\nu}$=0.96/41 dof. 

\noindent{\bf Bo 86} --  Leaving the column density as a free parameter, a power law
model provided a good fit with $\chi^2_{\nu}$=1.05/22 dof. The
best-fit parameters were $N_H=(1.28\pm0.50)\times 10^{21}$ cm$^{-2}$
and $\alpha=1.37\pm0.15$. The luminosity in 0.3--7 keV is
$5.26\times10^{37}$ erg s$^{-1}$. We have fixed the absorption at
$N_H=4.4\times 10^{20}$ cm$^{-2}$ [$E(B-V)=0.08$] and we obtained
$\alpha=1.15\pm0.08$ with $\chi^2_{\nu}$=1.15/23 dof. 

\noindent{\bf Bo 143} -- A Power law is the only model which can be fitted for the
spectrum. We have obtained $N_H=(1.96\pm0.66)\times 10^{21}$ cm$^{-1}$
and $\alpha=1.90\pm0.23$, with $\chi^2_{\nu}$=0.69/12 dof. The
luminosity in 0.3--7 keV is $3.17\times10^{37}$ erg s$^{-1}$. By
fixing the absorption at $N_H=7.2\times 10^{20}$ cm$^{-2}$
[$E(B-V)=0.13$], the best fit parameters were $\alpha=1.50\pm0.10$ and
$\chi^2_{\nu}$=1.02/13 dof. 

Except for the second observation of Bo 82, the power-law photon index
$\alpha$ ranges from 1.4 to 2 which is typical for low-mass X-ray
binaries. For Bo 82,  the photon
index changed from 1.4 to 1.1, indicating a transition from a soft
state to hard state but at a similar intensity level.  Trinchieri et
al. (1999) also found that Bo 82 to be a hard source and on that
basis conjectured that it might be a black hole. \rosat\ also observed
Bo 82 (see Supper \etal\ 1997); we reanalyzed the archival data and a power-law with
$\alpha\sim1.4$ gave a good fit to the data. 

We have constructed the time history of these four sources over $\sim
500$ days (see Figure 7). Just as for Bo\,375, Bo 82 and Bo 86 show intensity
variability by a factor of $\sim 3$. In particular, Bo 86 may have
long-term variability on timescales of $\sim 200$ days. Bo 153 and Bo
143 may be variable at a lower level but higher signal-to-noise light curves
would be needed to establish variability.

\begin{figure*}[t]
\begin{center}
\psfig{file=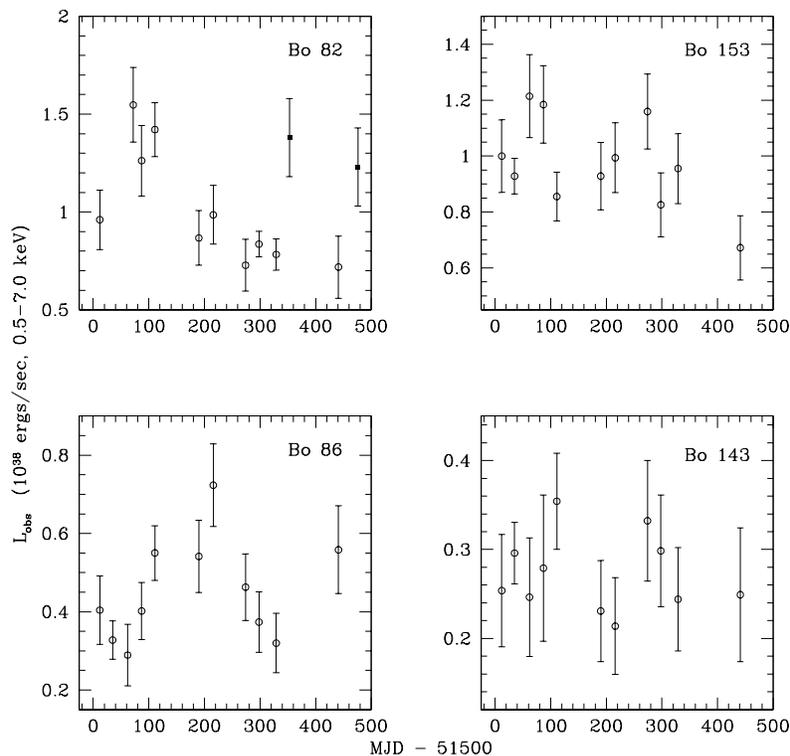,height=11cm,width=11cm}
\end{center}
\caption{\chandra\ HRC light curves of Bo 82, Bo 153, Bo 86 and Bo
143. The two ACIS-S pointings of Bo 82 are also included in the plot (solid squares).
The luminosity was estimated by the best-fitting spectrum which derived from
ACIS pointings. To compute its luminosity, we assumed
Bo 82 to be in its soft state 
($\alpha=1.4$).}
\end{figure*}

\section{Population Studies}

\subsection{The Luminosity Function}

Two types of information can be used to construct a luminosity function.
First, of  course, is the observed luminosities of the sources.
This is the information we present in this paper,
because our primary motivation is to feature the high-luminosity
sources that have actually been detected and to compare the detected
GC X-ray sources in M31 with detected GC X-ray sources in the Milky Way.
It is also possible to take a more comprehensive view that 
takes into account the
fact that we learn something about the luminosity function
from computing upper limits for clusters which have been observed
with X-ray detectors but within which no X-ray sources were discovered.
Ongoing and possible new observations may extend the depth of M31 
observations and their sensitivity to transient GC X-ray sources. Thus, 
a more complete picture of the M31 \gc\ X-ray sources is developing,
and will warrant a later careful look at non-detections as well as detections.

In Figure 8 we present the luminosity distribution of all $30$ 
GC X-ray sources observed by \chandra. Luminosities were derived
assuming a power-law model with $N_H=10^{21}$ cm$^{-2}$ and
$\alpha=1.7$ in 0.3--7 keV. The derived luminosities are not very sensitive to
the spectral parameters; the difference is about 20\% when
varying the $N_H$ from $6-15\times10^{-22}$ cm$^{-2}$ and $\alpha$
from 1.2 to 2. The thermal Bremsstralung model used
by Primini, Forman, \& Jones (1993) gave luminosity differences up to 80\%.

This luminosity function has $2$ quantifiable differences with the Galactic 
GC X-ray luminosity function, even though the
most recently published M31 GC luminosity  function found the M31
and Galactic GC X-ray luminosity functions to be in agreement (Supper
et al. 1997).
First, the peak X-ray luminosity
is higher by a factor of $~\sim 10$, 
and second, a larger fraction of all GC sources have luminosities
above $10^{37}$ erg s$^{-1}$ ($\sim 10/30$ in M31, vs $1/12$ in the Galaxy).

\begin{figure*}
\begin{center}
\psfig{file=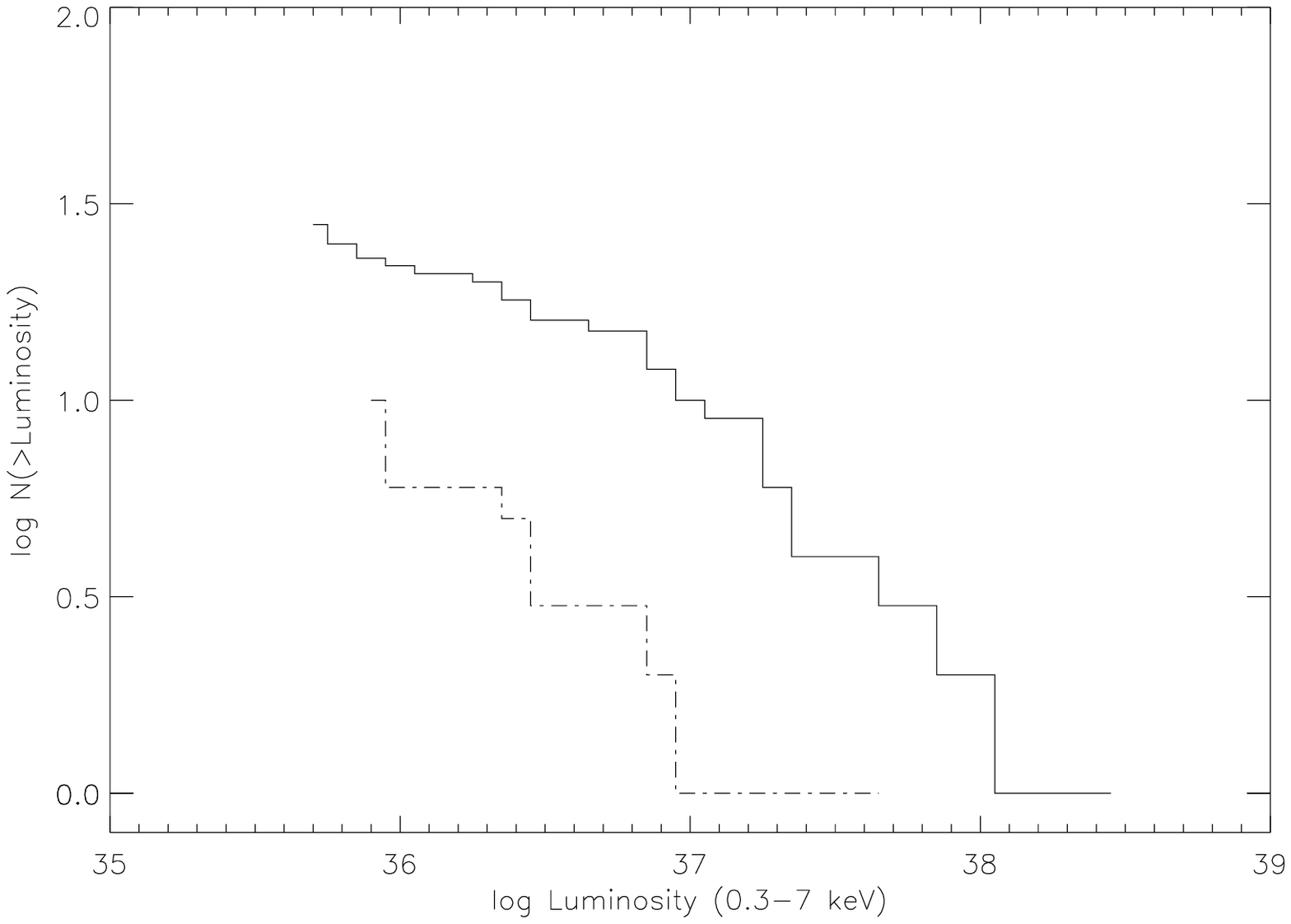,height=9cm,width=9cm}
\end{center}
\caption{Luminosity functions for M31 globular clusters and Galactic
clusters. The solid curve represents the M31 clusters and the dashed
curve is the distribution of the Galactic sources (Verbunt et al. 1995).}
\end{figure*}

\subsection{Previous Work}
 
The discovery of high-luminosity
sources in M31 X-ray sources began roughly $25$ years ago.  
Twenty one X-ray sources discovered by instruments aboard the Einstein
observatory were tentatively identified with globular clusters.
The source identified with the cluster labeled by Sargent
et al. (1977) as 307, which we have referred to as Bo\,375,
 had a measured X-ray luminosity of 
$2.7 \times 10^{38}$ ergs s$^{-1}$. The source identified with Bo 135
had a measured X-ray luminosity of 
$1.1 \times 10^{38}$ ergs s$^{-1}$. Long \& van Speybroek (1982),
noted that these luminosities were higher than any observed for
Galactic GCs.  Battistini \etal\ (1987) further noted that 
the fraction of then-known M31 GCs with X-ray sources might also be
larger than the fraction our Galaxy. This was because, even though the measured
fractions were similar, only the most luminous of the Galactic sources
could have been detected in M31 by Einstein.
This raised the question of whether the frequency of X-ray sources
in M31 was generally higher, or whether the GC X-ray sources in M31
were brighter than Galactic GC sources.
Observations of the central$\sim 34'$ of M31 with {\it ROSAT}'s HRI
discovered $18$ GC X-ray sources and placed
upper limits on  X-ray detection from $32$ additional globular clusters
(Primini, Forman, \& Jones 1993).
All of the measured X-ray luminosities were below
$10^{38}$ ergs s$^{-1}$, but $8$ were above $10^{37}$ ergs s$^{-1}$. 
A statistical analysis, based on survival analysis techniques
(Avni et al. 1980, Feigelson \& Nelson 1985, Schmitt 1985)
allowed  the upper limits as well as the measured luminosities to  be used
to construct the luminosity function. This procedure     
produced a cumulative X-ray luminosity distribution that 
exhibited a 
maximum luminosity consistent with that of
the  Galactic GC X-ray luminosity distribution,
but which seemed to have a larger population of high-luminosity
sources. The most comprehensive survey to date is the 
1991 {\it ROSAT} survey of M31, which detected $31$ \gc\ \xrs s (Supper et al.
1997; see also Supper et al. 2001). A statistical
analysis that paralleled the one carried out  by Primini et al. (1993)
agreed  that the maximum luminosity of M31 \gc\ sources
is consistent with that of
that of Galactic \gc\ sources. Based on this larger sample,
however, it was concluded that there is no difference between the 
\luf s of M31 and Milky Way \gc s.

Given the shifts in perception of the M31 GC luminosity function,
we note that the 
higher-X-ray luminosities ($>  10^{38}$ erg s$^{-1}$) are well
established in the {\it Chandra} data sets and in prior X-ray data.
Overall, using PIMMS to convert measured count rates to 
X-ray luminosities for the power-law model 
used to construct the {\it Chandra}-measured luminosity function,
we find that there are $\sim 7$ sources across data sets
with X-ray luminosities near or above $10^{38}$ erg s$^{-1}$.

Even the result about the  large fraction
of sources with $L_X > 10^{37}$ erg s$^{-1}$ is 
likely to remain
valid when additional observations of M31 GCs are carried out.
To justify this statement, we first note that
already-completed observations taken
by other X-ray telescopes 
can be used to extend the {\it Chandra} sample, and that the extended
samples exhibit this same characteristic. For example, 
the {\it ROSAT} data on $18$ GC X-ray sources (Supper et al. 1997, 2001)  
that are not in our fields includes $6$ GCs with $L_X$[0.3-7 \, keV] between
$10^{36}$ and $10^{37}$ erg s$^{-1}$, and $12$ GCs with
$L_X$[0.3-7 \, keV]$~> 10^{37}$ erg s$^{-1}$. 
Second, we note that the only way for the fraction of 
sources with $L_X$[0.3-7 \, keV] to be reduced to the Galactic value
would be if a very large number of M31 GCs were discovered
to house lower-luminosity bright sources ($10^{35} < L_X < 10^{37}$ erg s$^{-1}$).
The depth and spatial coverage of completed and ongoing
surveys seem to rule out this possibility.    
Furthermore, to discover enough 
lower-luminosity bright sources 
($L_X > 10^{35}$ erg s$^{-1}$) to bring 
the fraction of $L_X$[0.3-7 \, keV]$~> 10^{37}$
ergs s$^{-1}$ sources into line with the Galactic value ($1/12$), we would have to
find $\sim 200$ such sources. In this case, the fraction of M31 GCs with
X-ray sources would lie somewhere between $1/2-1$. The contrast
between this value and the value of $\sim 1/10$ for 
Galactic GCs which house bright X-ray sources would itself prove interesting.

\subsection{Spectral Comparisons}

Since our data and analysis show that there are clear differences in
the luminosity functions of the Galactic and M31 globular cluster
X-ray sources, it is important to compare the source spectra as
well. Because of the greater distance of M31, we can at present only
compare the spectra of its most luminous clusters with the spectra of
Galactic GC sources. In addition, the M31 GC X-ray spectra collected
so far, even for these bright sources, generally have fewer counts,
making direct comparison with the best Galactic GC data
difficult. Furthermore, if some of the bright M31 GC sources we
observe are actually composites of separate sources, blending may
further complicate the comparisons. Despite these caveats, it is
important to make a first step toward the systematic comparison of
spectra. Future observations which can provide better spectral and
spatial resolution will allow more detailed and meaningful comparisons
to the model.

\subsubsection{M31}

We used the fits to the ACIS-S observations described in \S 3. All
sources, except Bo\,82, as seen in an ACIS-S observation on 2001 March
8, can be fitted by a single power law model with $\alpha=1.4-1.9$. The
Bo\,82  spectra exhibited a somewhat harder ($\alpha=1.1$)
spectrum. Previous {\it BeppoSAX} observations provide another important
resource. Trinchieri \etal\ (1999) showed
that 8 of the M31 GC candidates spectra can be fitted with either a
power law ($\alpha=0.8-1.9$) or a bremsstrahlung ($kT=6-9$ keV) model
in 1.8--10 keV. Two of them (Bo\,82 and Bo\,158) have somewhat harder
spectra with $\alpha \sim 1$. Both the spectral fit of our Bo\,82
data (\S 3) and the binned count rates shown in Table 4, 
demonstrate that
these two sources are relatively harder than the
others. Therefore the {\it BeppoSAX} observations appear to be
consistent with ours, although source confusion could complicate 
the comparison.

\subsubsection{Comparisons with the Milky Way}
Callanan et
al. (1995) found from archival {\it EXOSAT} data that 5 out of the
6 Galactic GCs can be well fitted with a power law plus blackbody
spectral model ($\alpha=1.6-2$, $kT=0.5-1.5$ keV). Our high
signal-to-noise Bo\,375 spectra are consistent with their
results. 

In a recent study of GCs in our Galaxy using {\it BeppoSAX}, Sidoli
\etal\ (2001) showed that the X-ray spectra of GCs can extend up to
20--100 keV and they can be fitted with a black-body (or disk black-body)
plus Comptonization model. We here found the M31 GC sources also
have hard X-ray tails up to $\sim 10$ keV (which is the limit of both \asca\
and \chandra).
% an Comptonization component is required to provide a
%better fit. Although we only used power law to approximate the
%Comptonization component, it provides a hint that the cut-off energy is
%higher than 10 keV for the cut-off power law, or unsaturated
%Comptonization (Christian \& Swank 1997). 
We have also tried to fit
the spectra of \bol\ with COMPTT and COMPST models used by Christian \& Swank
(1997) and Sidoli \etal\ (2001); the parameters were not constrained
and the Comtonizing plasma temperature ($kT_e$) has a very large value ($\sim
100$ keV). This indicates that $kT_e$ is above 10 keV, which is also seen
in GCs in our Galaxy (e.g., Sidoli \etal\ 2001).

To facilitate the comparison between M31 sources and Galactic
sources, we reanalyzed all the archival {\it BeppoSAX}
Galactic GCs used by Sidoli et al. (2001). The reason for using {\it BeppoSAX}
data is that it covers more samples and has a common energy range with
{\it Chandra}. We used an absorbed power law model (with $N_H$ fixing
at the optical determined value). We also 
 restricted the fit to energies ranging
from 2--7 keV for both data sets. The reason for this is that
absorption plays a significant role in shaping the observed spectrum from
some of the Galactic sources; we wished to compare the Galactic
and M31 sources across a range of energies in which absorption is not
important.  
Table 5 shows the results of the {\it BeppoSAX} spectral fits together
with the {\it Chandra} fits on M31 sources in the same energy
range. 
Although the fits are unacceptable for some cases, and are
not expected to correspond to the true physical models,
due to our limited energy range, the derived
$\alpha$ is in the range of 1.4--2.5, while M31 sources have values
between 1.4 and 1.9. These results are in good agreement, 
although the sample size of M31 sources is
limited. We also used PIMMS to estimate the {\it Chandra} count rate
for a single power law spectrum with $\alpha=1.4-2$ and $N_H=10^{21}$
cm$^{-2}$ and calculated the hardness ratio (2--10 keV/0.5--1.5
keV). The hardness ratio ranges from 0.4 to 0.8 which is consistent
with most of the values shown in Table 4. Combining the M31 GC
sources observed with {\it BeppoSAX} in 1.8--10 keV (Trinchieri et
al. 1999), we conclude that the X-ray spectral properties of M31 GCs
and Galactic GCs above 2 keV are roughly the same.

\begin{table*}
\begin{center}
TABLE~5

{\sc Best-fit Spectrum (2--7 keV) of Galactic and M31 Globular
Clusters}

\vspace{1mm}
\begin{tabular}{lcccc}
\tableline \tableline
\noalign{\smallskip}

Source & Distance &$N_H$ & $\alpha$&$\chi^2_{\nu}/dof$\\
       & (kpc)& $(10^{21}$ cm$^{-2})$&\\ 
\hline
NGC1851 & 12.2&0.18&$1.98\pm0.05$&1.3/51\\
Terzan 2& 10.0&7.16&$1.44\pm0.04$&15.9/51\\
NGC6441 & 10.7&2.32&$1.63\pm0.05$&5.4/51\\
Terzan 6& 7.0&13.24&$1.82\pm0.01$&1.0/51\\
NGC6624 & 8.1&1.61&$1.93\pm0.03$&3.2/51\\
NGC6712 & 6.8&2.51&$2.46\pm0.01$&1.1/51\\
NGC7078 & 10.5&0.36&$1.85\pm0.09$&8.3/50\\
NGC6440 & 7.0&6.27&$1.62\pm0.01$&1.2/51\\
Terzan 1& 5.2&12.17&$2.00\pm0.15$&1.5/28\\
&&&&\\
Bo\,375 & 780& 0.61 & $1.86\pm0.05$&0.96/115\\
Bo\,82 (1st)& 780 & 4.00 & $1.67\pm0.09$&0.81/57\\
Bo\,82 (2nd)& 780    & 4.00 & $1.32\pm0.10$&0.97/47\\
Bo\,153& 780& 0.60 & $1.53\pm0.19$&1.02/14\\
Bo\,86& 780 & 0.44 & $1.43\pm0.30$&0.32/6\\
Bo\,143& 780 & 0.72 &$1.89\pm0.75$&0.37/1\\
\noalign{\smallskip}
\tableline\\
\multicolumn{5}{l}{NOTES --- see \S 4.3.2 for discussion.}\\
\end{tabular}
\end{center}
\end{table*}

These comparisons make it clear that, in spite of significant luminosity
differences, the conjecture that the M31 \gc\ X-ray sources have
spectra similar to those of the Galactic GC sources is still viable. Nevertheless these broad band
comparisons are not able to either establish a neutron star nature or
rule out a black hole nature for the sources. The need for further
observations is emphasized by the fact that observations of Bo 82 have
found it to be sometimes in a luminous soft state, similar to Galactic GC
sources, and sometimes in a luminous hard state.

\section{The M31 Globular Cluster System vs the Milky Way's Clusters}

The fact that X-ray binaries are as luminous as the ones we have
studied in M31 \gc s is not problematic. These sources do present a
puzzle, however, when we compare them to \gc\ X-ray sources in our
own Galaxy. 
The primary differences between the population of X-ray sources in
M31 \gc s and those in Galactic \gc s are that (1) the peak
luminosity is higher, and (2) the high-luminosity end of the
distribution function is more populated.
It is natural to ask if these differences can be explained by differences in
the two galaxies' populations of \gc s.

There is one obvious difference between the \gc\ systems of M31
and the Milky Way: M31 has more \gc s. With $\sim 150$ well documented
Galactic \gc s, the total population could be larger by $\sim 20$
(Minniti 1995). The total number of M31 \gc s is less certain,
but there are over $200$ confirmed clusters (Barmby et al.\ 2000)
and the total population could be $\gtrsim 400$ (Battistini et al. 1993).
On a practical level, the greater distance to clusters in M31
has made it difficult to learn as much about individual clusters
as we know about Galactic \gc s.
HST has enabled studies of M31 \gc\ morphology
(Bendinelli et al. 1993) and stellar populations (Ajhar et al. 1996),
although not at the same level ground-based studies have achieved for
\gc s in our own Galaxy. HST studies, which have concentrated on the
brightest M31 \gc s, find no obvious differences between the properties
of these clusters and those of the brightest Milky Way clusters.
Ground-based observations of M31 \gc s have provided a wealth 
of information about integrated luminosities, colors, and 
spectroscopic features of the population as a whole. The result
of these studies is that the \gc s of M31 are similar in most respects to
those of the Galaxy. Differences tend to be subtle: for example,
M31 GCs are overabundant in CN compared to Milky Way GCs of
the same metallicity (e.g., Burstein et al. 1984).
The M31 globular cluster luminosity function also shows evidence for
variation not seen in the Milky Way GC luminosity function: 
the inner clusters are
on average, brighter than the outer clusters, and the metal-rich clusters
are brighter than the metal-poor clusters (Barmby, Huchra, \& Brodie 2001).
Assuming the GC luminosity function differences are real and not
caused by catalog incompleteness,
they could point to age or mass variations in the M31 GCs (see \S 6).

\subsection{Fraction of Clusters with X-ray sources: Radial Dependence}

Since we have imaged four fields at different locations in
M31, we can examine the fraction of clusters with X-ray sources
as a function of projected distance from the center of the
galaxy. To do this, we used the GC catalogs 
to determine which clusters fell into the region of sky 
covered by each image. The number of X-ray detected clusters divided by
the total number of clusters in each image is the fraction of
clusters with X-ray sources.

Determining this fraction is not as straightforward as it seems:
while the number of X-ray detected objects is well-understood,
the number and identity of all true globular clusters in
M31 is not. 
The true number of clusters in the central regions is particularly
uncertain: \gc s are difficult to detect against the bright,
variable stellar background, and the existing catalogs may be incomplete.
We therefore computed two fractions: the fraction of
confirmed globular clusters detected as X-ray sources, and the 
fraction of all GC candidates detected.  
Since only a fraction of candidates
are confirmed, and not all cluster candidates are true clusters,
these two values should bracket the true fraction of M31 clusters
with X-ray sources.

Figure~9 shows these fractions for each field as a function of
the projected distance, $R_p$, from the center of M31. 
The figure also shows the fraction of Milky Way
clusters with X-ray sources in four bins of $R_p$,
where $R_p$ was computed for each cluster as if the Milky Way globular cluster
system were viewed from the same inclination angle as the M31 GCS.
Although in M31 we have only four fields and 30 detected clusters, 
there is a clear decline in the fraction of M31 clusters with X-ray sources 
as $R_p$ increases, and a suggestion of the same effect in the Milky Way.
The Milky Way curve is bracketed by the two M31 curves
(the dip at the second point is probably not significant
given the small number of Galactic X-ray sources),
suggesting that the pattern in the two galaxies is similar.
% is this the first time this effect has been demonstrated? say so.
The decline in the number of X-ray sources as $R_p$ increases
can be understood if dynamical evolution drives the creation of 
X-ray binaries, since dynamical evolution is expected to
be accelerated for clusters with orbits that bring them closer to the 
center of a galaxy.

\begin{figure*}[t]
\begin{center}
\psfig{file=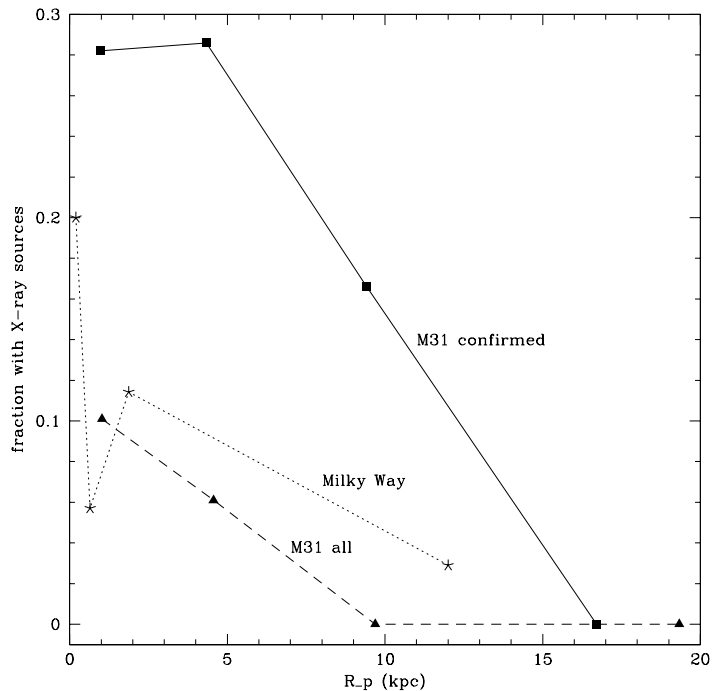,height=10cm,width=10cm}
\end{center}
\caption{Fraction of GCs with bright X-ray sources as a function of
projected separation, $R_p,$ from the galaxy center.}
\end{figure*}

The overall fraction of clusters with X-ray sources is fairly similar 
in the two galaxies. The Milky Way has 12 X-ray source clusters
and a total of 147 clusters, so $\sim 10\%$  of clusters have X-ray sources.
In our M31 fields, the fraction of confirmed clusters with
X-ray sources is 25\%, and the fraction of all cluster candidates
with X-ray sources is $\sim 7\%$. Again, to within our admittedly large uncertainties,
the fraction of clusters with X-ray sources seems to be fairly similar 
in the two galaxies.
It is therefore tempting to view the combined Galactic
and M31 globular clusters systems as a single population of clusters.
In this view, the clusters in M31 can help us to learn more about
conditions in Galactic globular clusters,
mainly by providing a larger population of clusters to study.

\subsection{Properties of Clusters with X-ray Sources}

Do M31 GCs with X-ray sources differ from those without? We compared 
optical colors and apparent magnitudes, color excesses, metallicities, 
radial velocities, and positions of the two populations using data 
drawn from the Barmby et al.\ (2000) catalog. We also compared
estimates of the core radii, from Crampton et al.\ (1985),
and estimates of the clusters' ellipticities, from
Staneva et al.\ (1996) and D'Onofrio et al.\ (1994)
(these estimates are from ground-based optical imaging and are not
particularly precise; their major advantage is that they are available
for most of the clusters). A KS test was used to determine
whether the two populations were drawn from the same distribution.
The only significant difference between the X-ray source and non-X-ray-source
clusters is in luminosity.
The clusters with X-ray sources are brighter
than the non-X-ray clusters, at the 99\% confidence level.
The luminosity difference of the median values is
0.55 magnitudes, or a flux difference of a factor of 1.7. 

For 20 clusters (6 with X-ray sources and 14 without), structural parameters 
were also available from HST images (Barmby \& Holland 2001). We compared the
properties of the two groups and found only one significant difference: the central 
surface brightness of the X-ray source clusters was much brighter (median ${\mu}_0,V=14.7$) 
than that of the non-X-ray-source clusters (median ${\mu}_0,V=16.3$). This result is
only significant at the 92\% level and is based on a small number of objects.
However, it is consistent with the difference in optical luminosities found above,
as studies of larger populations of Milky Way and M31 clusters show that more
luminous clusters tend to have higher central surface brightnesses (Djorgovski \& Meylan
1994; Barmby \& Holland 2001). In Figure 10 we show all the available structural properties 
for M31 and Milky Way clusters versus projected distance from the center of the galaxy.
The figure shows that the X-ray source clusters tend to be near the center of their parent 
galaxies, and as such tend to have high central surface brightnesses, small
scale and half-light radii, and high concentrations. 

\begin{figure*}[t]
\begin{center}
\psfig{file=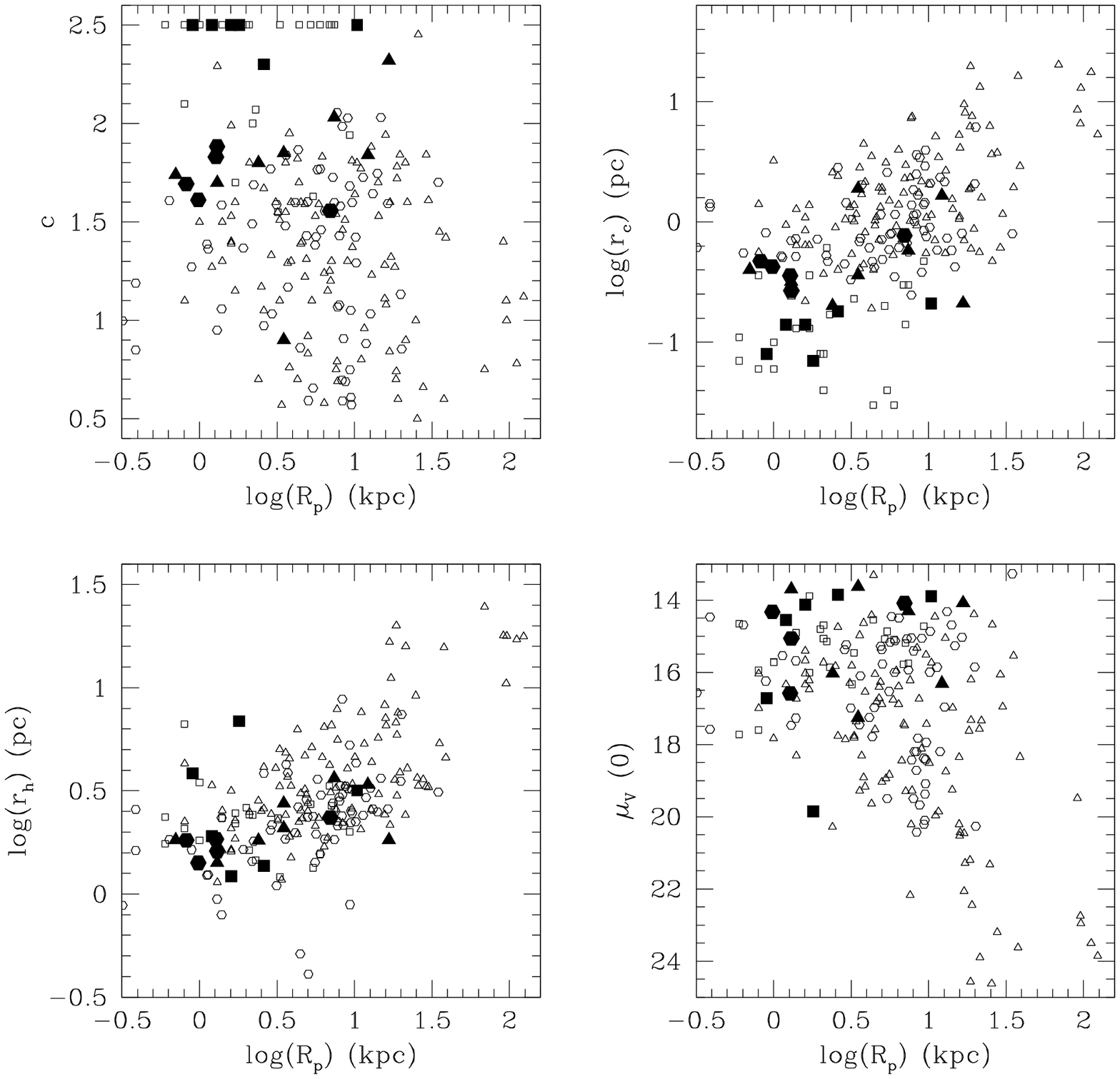,height=10cm,width=10cm}
\end{center}
\caption{Structural parameters versus projected distance $R_p$ from the galaxy center for
M31 and Milky Way globular clusters. Core-collapsed Milky Way GCs are squares, 
non-core-collapsed Milky Way GCs are triangles, and M31 GCs are hexagons. 
(Definitive detection of core-collapse in M31 GCs is not possible with
existing data). Filled symbols represent clusters with bright X-ray sources;
open symbols represent all other clusters. Globular cluster
surface brightness distributions are described by King models with
three parameters: $r_0$, the scale radius (sometimes mis-called the core radius), 
$c$, the concentration ($c=\log(r_t/r_0)$, where $r_t$ is the tidal radius), and
${\mu}_0,V$, the central surface brightness. $r_h$, the half-light radius,
is another commonly-used measure of cluster size which combines $c$ and $r_0$.
Data for the Milky Way clusters is from the Harris (1996) catalog;
data for the M31 clusters is from Barmby \& Holland (2001), which contains
a detailed comparison of M31 and Milky Way clusters' structural properties.}
\end{figure*}

We also compared the properties of M31 clusters with X-ray sources
to those of Milky Way clusters with X-ray sources. The only possibly
significant difference was in the luminosity: the M31 X-ray clusters
were brighter in $V$ than the Milky Way X-ray clusters. This is not true of
the GC populations in general --- the overall globular cluster luminosity
functions are not significantly different --- so it could be a clue to
the M31 GCs' higher X-ray luminosities. It would be interesting to know 
whether the higher optical luminosities of the M31 X-ray GCs translate
into higher masses; unfortunately, mass determinations from velocity dispersions
are available only for a few M31 GCs, and only one of the X-ray sources.
There are no clear correlations between M31 GC X-ray source luminosity and
any of the cluster properties mentioned above. 

\section{Conclusion: Possible Explanations}

\subsection{Individual Bright M31 GC Sources}

Some individual X-ray sources in M31 globular clusters are
significantly more luminous
than any individual X-ray source yet observed in Galactic globular clusters.
Bo\,375 is an example. With
changes in flux at the $\sim 50\%$
level over $16$ hours and at the $100\%$ level over times ranging from
months to years, it seems clear that a single component must be contributing
at least half of the X-ray energy emitted at typical times.
This source would, by itself, approach the Eddington limit for a $1.4\, M_\odot$
neutron star, and must be more than twice as X-ray luminous as
the maximum $L_X$ measured for 
any Galactic globular cluster X-ray source.
We find similar variability effects in the 500-day light curves of
$2$ other highly luminous M31 GC X-ray sources,
Bo\,82 and Bo\,86. It may therefore be the case that the most luminous 
sources we observe in M31 GCs 
are either individual X-ray binaries 
(although there is  evidence in the ACIS-S image that Bo\,375 may not be
a point source), or composites of a small number of independent
X-ray binaries.  
An alternative explanation might apply to some of the
sources not yet known to exhibit significant short-term variation:
that some of the more 
luminous sources ($L_X > 5 \times 10^{37}$ ergs s$^{-1}$)
are
composed of a fairly large number ($\sim 10$) of typical
Galactic globular cluster X-ray binaries. 
Although not ruled out, such a large
number of sources in roughly $1/4$ of M31's globular clusters, with
no composites at all yet observed in Galactic globular clusters,
seems extreme (see \S 6.2).

It is clear that the physical characteristics of
 at least those very bright M31 GC 
sources exhibiting significant short-term variability
differ from those of Galactic GC sources.
We have therefore analyzed the spectra of the $5$ brightest M31 sources 
(the only M31 GC sources for which we have collected more than $400$ counts)
and compared them, and also M31 GC X-ray 
 sources studied previously with {\it BeppoSAX}
(Trinchieri \etal\ 1999),  with the spectra of Galactic sources.
The spectra are similar to those of Galactic GC
sources, but $2$ sources exhibit somewhat harder power-law tails; one
 of these appears to make transitions between a soft and a hard state. 
It is not possible  to draw conclusions about the
physical nature of the sources from these crude comparisons (which
are nevertheless, the best presently possible). We must therefore
continue to carry out time variability and spectral studies that
could possibly, in combination, tell us whether each source
is an accreting black hole or neutron star.
High-spatial resolution observations with these sources on-axis,
and spectra from longer observations with {\it Chandra} and {\it BeppoSAX}
could play important roles, as could {\it XMM-Newton} and HST observations. 
Progress along theoretical lines could also
help to 
distinguish signatures of a relatively steady
black hole accretor from those of a neutron star
accretor.

If it is an accreting black hole, 
the time signatures of the source in Bo\,375 are more
characteristic of a persistent, rather than a transient source.
Although this would at first suggest a high-mass ($M>5\, M_\odot$) donor,
such  a donor is unlikely to be found in a globular cluster--even if
the cluster is relatively young and even if the rate of stellar intersactions
near the cluster core is high.  
Nevertheless, if a low-mass donor is evolving away from the main sequence,
a black hole accretor that 
might otherwise be transient could be highly
luminous over relatively long times.

Below we construct a specific neutron star model for Bo\,375 
suggested by the 16 hour time
scale observed by {\it ROSAT}. This model may or may not describe the
actual physics of Bo\,375, but it  is consistent with the data
and could apply to other bright GC sources in the youngest clusters
of M31 and other galaxies.

\subsection{A Model for Bo\,375}

The spectral and timing studies we have been able to carry out
so far have not provided promising clues as to the nature of this X-ray 
source (or sources). It is intriguing, however, that 
{\it ROSAT} observations found what appeared to be systematic,
possibly periodic, variations on a time scale of $\sim 16$ hours.
If these variations are associated with an orbital period,
this would suggest that mass transfer is occurring on the thermal-time-scale
of a slightly evolved Roche-lobe filling star.

\subsubsection{Thermal Time Scale Mass Transfer}

Thermal-time-scale mass transfer is driven by the reaction of 
a Roche-lobe-filling donor
star to mass loss (van den Heuvel et al. 1992; Kalogera \& Webbink 1996; Di\,Stefano \& Nelson
1996; King et al. 2001; King \& Begelman 1999). 
Consider  a donor star which would, were it living in
isolation, have an equilibrium radius $R_{eq}$ determined by
its total mass and state of evolution.
If this star is in a close binary and is filling its Roche lobe,
in a situation in which $R_L$ is consistently somewhat
smaller than $R_{eq}$, 
then the drive of the star to achieve equilibrium causes
it to lose mass through the L1 point at a relatively rapid rate.
In fact, if the ratio of the donor mass  to the accretor mass is
too large, or if the donor is evolved enough to be fully
convective, the mass transfer process can become dynamically unstable,
leading to the formation of a common
envelope.  
If, however, a dynamical instability is avoided, mass transfer is
driven on a time scale closely related to the thermal time scale
of the  donor.                              

In general, the two assumptions of a Roche-lobe filling donor and
angular momentum conservation lead to an analytic expression for 
$\dot m,$ the rate at which the donor loses mass. This expression
can be written generically as
\begin{equation} 
\dot m = {{{\cal N}}\over{{\cal D}}},
\end{equation}
where ${\cal N}$ and ${\cal D}$ are each generally functions of
the masses of the donor and accretor;
the state of evolution of the donor, which can be parameterized by its 
core mass, $m_c$; $\beta,$ the fraction of
the mass incident on the accretor that is actually retained;  
parameters that quantify the amount of angular momentum dissipated 
or else carried away by matter ejected from the system;
and the orbital separation $a$ between the components of the
binary. The value of ${\cal D}$ determines whether mass transfer can
be dynamically stable.

The numerator is a sum of terms: 
${\cal N}_w$ is related to the donor's stellar winds; 
${\cal N}_{gr}$ and ${\cal N}_{mb}$ are governed by 
gravitational radiation and  
 magnetic braking, respectively;  
${\cal N}_{nuc}$ relates to the nuclear evolution of the donor; 
${\cal N}_{th}$ relates to the thermal-time-scale readjustment
of the donor to the size of the Roche lobe.  
Different regions  in the parameter space correspond to
different physical systems. For example, mass transfer in systems with
large values of $m_c$ and $a$ is
driven by the nuclear evolution of the system; ${\cal N}_{nuc}$ is
the dominant term.  

${\cal N}_{th}$ can be the dominant term when the donor's mass
is comparable to or greater than that of the accretor, especially when the
donor is somewhat evolved, but does not yet have a fully convective
envelope (typically $0.05 M_\odot < m_c < 0.2 M_\odot$).  
Orbital periods are typically on the order of a day. 
In such cases, the rate of mass transfer can be expressed as
\begin{equation} 
\dot m \sim \Bigg({{R_{eq}-R_L}\over{R_{eq}}}\Bigg) 
            \Bigg({{1}\over{\tau_{KH}}}\Bigg)\, 
            {\cal F}_{dyn}\, {\cal G}_{star}   
\end{equation} 
In the expression above, $\tau_{KH}$ is the Kelvin-Helmholz time of the
donor star ($\sim 10^7$ years for a star of approximately $M_\odot$); 
${\cal F}_{dyn}$ is a factor that depends on the mass ratio
of the stars and on the amount of angular momentum carried away by any
mass exiting the system or else lost through dissipative processes
like gravitational radiation; ${\cal G}_{star}$ expresses the donor's response
to mass loss; this factor must be computed  by a Henyey-like calculation.
The favored values for the calculations we present below are taken
from work by Nelson (1995). For more details about this type of mass transfer,
see \rd\ \& Nelson (1996) and references therein.  

\subsubsection{System Properties}

When the accretor is a neutron star, only a narrow range of donor masses
and states of evolution (as specified by the helium core mass)
will permit dynamically stable mass transfer driven on the thermal
time scale of the donor. In Figure 10  
we have shown the range of system parameters for
such systems at that point in their
evolution at which $P_{orb}$ is $16$ hours and the accretion
rate onto a neutron star companion has just increased 
to $10^{-8} M_\odot.$ Donor masses range from $1.1-1.6\, M_\odot$ and
donor core masses are near $0.11\, M_\odot$. These systems are similar
to the ``intermediate-mass binaries'' studied by Davies \& Hansen
(1998) in significant young GCs.

\subsubsection{Evolution}

Although we don't know that the $16$-hour time variability of Bo\,375 is
related to an orbital time scale, it is reasonable to entertain this 
possibility. To this end we computed the  evolution of
tens of thousands of systems. Shown in Figure 11 are the results  
of some typical evolutions of binaries that would have accretion energies
greater than $10^{38}$ ergs s$^{-1}$ during a time interval in which
their orbital periods are $16$ hours. Note that during the early parts of 
the evolution, the donor's radius (which is governed by its total mass
is also influenced by its core mass), and the orbital period can
decrease.

\begin{figure*}[t]
\begin{center}
{\rotatebox{-90}{\psfig{file=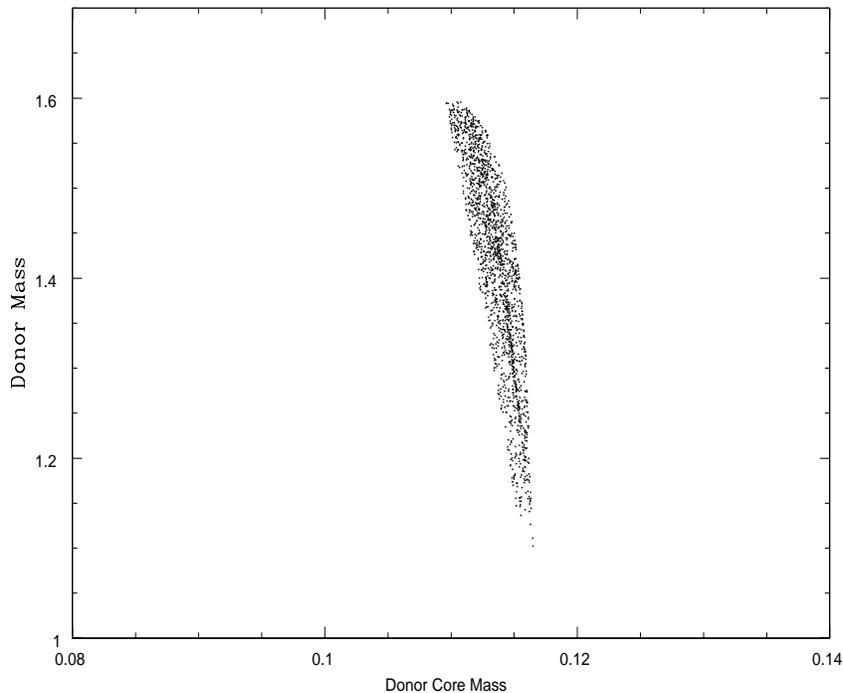,height=12cm,width=10cm}}}
\end{center}
\caption{Each dot corresponds to a physical system with an
orbital period of $16$ hours, and an
accretion luminosity above $10^{38}$ erg s$^{-1}$.}
\end{figure*}

\subsubsection{Viability in a Globular Cluster Environment}

This sort of scenario  has not been considered for neutron stars in GCs, because the 
required mass of the donor star is larger than the mass of stars at the 
turn-off of typical Galactic GCs (Di\,Stefano \& Davies 1996 have
considered such a scenario for white dwarfs; see Hertz, Grindlay, \&
Bailyn 1993; Verbunt et al. 1994, 1995). 
There are two sets of circumstances, however, that
alone or in combination can provide a pool of massive donor stars.
The first and simplest is a younger cluster age.
Although most Galactic GCs seem to have been formed between $12$ and $15$ Gyr
ago, other galaxies may have experienced several distinct epochs of globular
cluster formation. This seems to have been the case in the Magellanic
Clouds (see, e.g., Da Costa, Mould, \& Crawford, 1985),
and is also expected in galaxies that have experienced interactions
with neighboring galaxies. 
We do not know if M31 contains GCs formed during different epochs, but
 Barmby, Huchra, \& Brodie (2001) have suggested that
the difference in optical luminosity between the inner and outer GCs in M31
can be explained if the brighter clusters
are $\sim 55\%$ younger than the fainter clusters.
Young clusters formed $7-9$ Gyr ago could harbor slightly evolved stars that
would fit the criteria shown in 
Figure 10 for a thermal-time-scale 
mass transfer system with a $16$ hour period.

In addition, a relatively high probability of interactions enhances
the probability of finding a massive star in a close orbit 
with a neutron star. 
If, for example, interactions
among massive stars in the core region have led to stellar mergers,
thermal-time-scale mass transfer onto a neutron star could be expected
occur in GCs. 
 
\subsubsection{Tests}

One test of the applicability of our model to individual bright
sources in M31 would be the discovery of time $\sim 1$ day 
variations among the 
M31 GC sources (or in other galaxy's GCs) 
that can be reasonably linked to an orbital period.
Another test can be applied to galaxies with GC systems that house  
highly luminous sources that are candidates for our model:
 are the relative numbers of systems of different luminosities
consistent with the lifetimes of the corresponding  phases of binary 
evolution?
We find, for example, 
that the epoch during which the  luminosity remains above $10^{38}$
ergs s$^{-1}$ is approximately $10^7$ years, roughly $\tau_{KH}$ for the donor. 
The lifetime of the accretion luminosities above $10^{37}$ ergs s$^{-1}$ are 
$\sim 10$ times longer. This roughly matches the statistics
of the M31 GC sources.
Finally, as we probe the globular cluster systems of more 
distant galaxies, we can
search for correlations between the X-ray luminosities of GCs and the 
ages of GCs. While no individual test is likely to be definitive, a
combination of them could provide support for, or else falsity, the model.
 
\begin{figure*}[t]
\begin{center}
\psfig{file=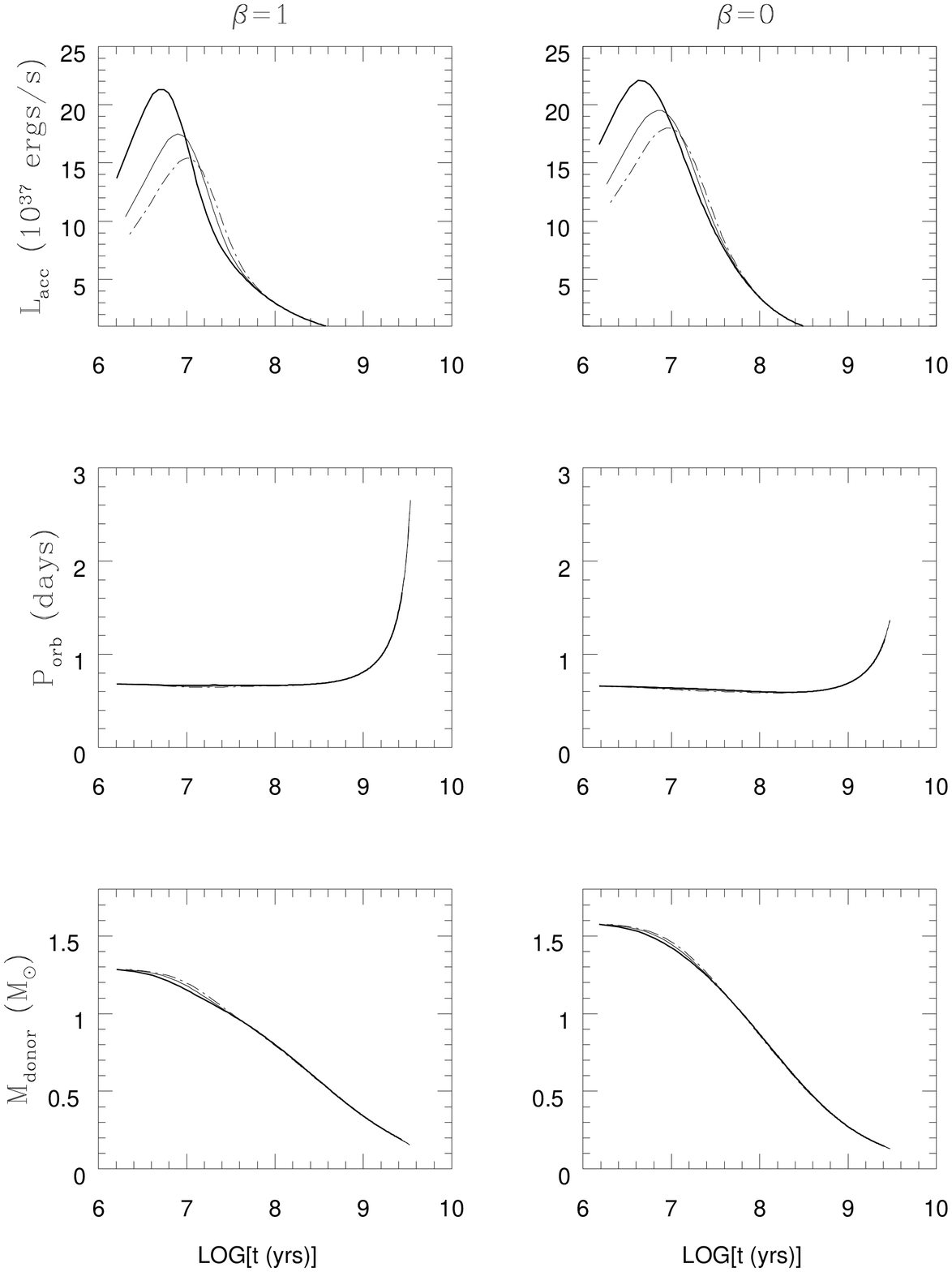,height=12cm,width=10cm}
\end{center}
\caption{Evolution of a typical system which has accretion luminosity above
$10^{38}$ erg s$^{-1}$ during a time when its orbital period is $16$
hours. $\beta = 1$: In this case, the neutron star accretes all of the
incident matter. If, however,
$\beta=1$ would produce super-Eddington accretion, the value of
$\beta$ is calculated in an internally-consistent way (see Di Stefano
2001). Note that for non-zero $\beta,$ the mass of the neutron
star can increase. $\beta = 0$: the mass of the neutron star is held
constant. In both cases the middle curve was computed using the values
for the thermal time-scale readjustment of the donor computed
by Nelson (1995). The upper and lower curves tested the effects of
altering these values (which are uncertain)  
by about an order of magnitude.}
\end{figure*}
 
\subsection{Other M31 Globular Clusters}

This paper started as a description of an interesting
X-ray luminous globular cluster, Bo\,375, which happened to
be in the field of view of a {\it Chandra} census of
M31. Its high luminosity, together with indications that the source
is not point-like, attracted our attention to this cluster.  
The fact that others of the most luminous sources studied in the 
census are also associated with globular clusters
led us to extend our investigation and to find that
the high-end of the the GC \xrs\ luminosity function has a higher
peak luminosity and a larger fraction of all sources populating the 
region with $L_X > 10^{37}$ ergs s$^{-1}$.  
The fundamental issue we address in this section, is
why M31's globular clusters should be more likely than those
of our own Galaxy to house brighter X-ray sources and
multiple X-ray sources. We find that three effects are likely to be at work.

\subsubsection{Possible Explanations: I. Large Population of Clusters}

Let us entertain the hypothesis that the Milky Way and Andromeda 
\gc s are members of a single distribution of clusters. 
In this case, the total population of clusters could be $\sim 550.$ 
We certainly wouldn't expect that every possible cluster phenomenon could be
discovered in the arbitrary subset of $150$ that we  happen to find in our
own Galaxy. If, e.g., black hole binaries are only $1/20$ as 
likely to form and remain in clusters as are neutron star binaries,
we would have a small chance of observing one among the  $14$ MW GCs
with \xrs s, but  would have an improved
 chance of
discovering one in the $\sim 50$ X-ray active GCs in M31. 
Unfortunately, however, we have no {\it a priori} estimates of
the probability of events (such as the formation of a black-hole
binary) that (a) have never been observed, and that (b)
are not readily amenable to reliable first-principles predictions. 
Thus, all that can be said is that we should perhaps not be too
surprised if some of M31's GCs do contain actively accreting black holes, or
unusually massive (e.g, $2-3\, M_\odot$) accreting neutron stars,
or even unusually massive donor stars.

There is, however, one yet-to-be-observed GC situation, for which
the probability can be estimated. This is the situation in which a cluster
houses more than one bright X-ray binary.
If $p$ represents the probability of finding an \xrs\ in a GC,
Poisson statistics predict
that, for every GC with a single \xrs , there should be $p/2$ clusters
with $2$ sources and $p^2/6$ clusters with $3$ sources. 
If $p \sim 0.1-0.2$, we would predict that $3-5$ GC \xrs s in M31 should 
be composites of $2$ independent sources, while fewer than $1$
is likely to be composed of $3$ independent sources.

One can also attack this question of the probability of multiplicity
from the perspective of the probability 
of the $2-$, $3-,$ and $4-$body processes that produce X-ray binaries.
What is the probability that two interactions
will produce $2$ independent \xrs s that are active at the same
time? Such a question certainly can be well posed and answered
subject to input assumptions about the lifetimes of X-ray binaries,
Since, however, current estimates of the lifetime of X-ray activity
vary by $1-2$ orders of magnitude, the most reliable answer
comes from the phenomenology that allows us to estimate $p$.
It is important to note, however, that if a subset of clusters have
different physical characteristics, the true probability of observing
multiple X-ray sources in these GCs could be different,

\subsubsection{Possible Explanations: II. More Highly Evolved Clusters}

Although there are striking similarities between the
MW and Andromeda galaxies,
they are not identical. In fact there are differences that are
likely to lead to more or (``enhanced'') accelerated evolution of GCs close to the center
of M31 than of clusters close to to the center
of the MW. According to van den Bergh (2000), the nuclear mass
of M31 is almost $30$ times larger than the nuclear mass of the MW
($7 \times 10^7 M_\odot$ vs $2.5 \times 10^6 M_\odot$). 
The bulge dispersions are $155$ km/s and $130$ km/s, respectively,
while the rotational velocities are, respectively, $260$ km/s vs $220$ km/s. 
All of these differences indicate that GCs close to the center of
M31 are likely to evolve more quickly than clusters close to the
center of our Galaxy (see, e.g., Gnedin, Lee, \& Ostriker 1999, and references therein). Since cluster evolution leads to smaller and denser
cores, at least until the onset of core collapse, the conditions
that lead to the formation of X-ray binaries may also be enhanced,
leading to more X-ray binaries in M31 GCs. 
Certainly the fact that there are more X-ray luminous clusters 
closer to the centers of both M31 and the MW speaks in favor of the 
role of accelerated evolution. 
Barmby, Huchra \& Brodie (2001) find an apparent dearth of low-mass
clusters near the center of M31, consistent with theoretical 
calculations of accelerated evolution; however, they find that the width of the
optical luminosity function is not consistent with the same calculations.

There is an argument that   
may not favor enhanced accelerated evolution as the primary reason for
the brighter X-ray luminosities of M31 GCs. This is that, so far, the
fraction of M31 \gc s in each region of the galaxy seem comparable to
the fraction observed at  similar values of $R_p$ for the MW.
As we have pointed out, 
these results are subject to significant uncertainties.  
But, if it is true that the fraction of GCs with X-ray sources 
is indeed   comparable for comparable values of $R_p,$ 
this would appear to limit the role of enhanced accelerated evolution.
It could, however, still be invoked to explain higher cluster X-ray luminosities
in M31 through the formation of more or brighter binaries if, e.g., 
the time scale for X-ray emission is only a short fraction
of the cluster's orbital period. 

In summary, it is difficult to quantify the role of accelerated evolution
due to interaction of M31's GCs with the rest of the galaxy.
Nevertheless, it is likely that it does play a somewhat different role
in M31 than in the MW, simply because of the differences in mass
and mass distribution in the $2$ galaxies. 
If, therefore, we wish to view the two systems of \gc s 
as a single larger system of
some $400-550$ clusters, we must invoke the  proviso that
none of the Galactic clusters experience tidal and disk interactions as
extreme as those experienced by clusters which pass close to the
center of M31. That is, if we were to map Galactic clusters
onto M31, in such a way that their interactions with M31 would
mimic the conditions each actually experiences due to the Milky Way,
the Galactic clusters would generally be placed farther from the
center of M31 than their actual distance from the center of the Galaxy.
Thus, M31 may be expected to contain not only \gc s with properties that
are similar to those found in our Galaxy, but also a sub-population
whose evolution is proceeding at a more rapid pace due to interactions
with M31; these more rapidly evolving clusters are most likely to be found
near the center of M31. Note that if enhanced accelerated evolution to
reponsible for the more luminous GC X-ray sources in M31, then M31 GCs
with bright X-ray sources should have different structure
parameters. HST observations could therefore help to determine the
role of enhanced accelerated evolution.

\subsubsection{Possible Explanations: III. Younger Clusters}

We have found that the M31 GCs with X-ray sources tend to be optically 
brighter than clusters without X-ray sources. Barmby et al.\ (2001)
suggested that a younger age might account for the higher average
luminosity of the metal-rich M31 GCs compared to the metal-poor clusters.
Metallicity is not strongly correlation with the probability that
an M31 GC has an X-ray source; however, it is interesting to
consider the possibility that the X-ray source clusters are
optically brighter because they are younger. If this is true,
the X-ray source clusters would have a higher turn-off mass
(1.1--1.2 $M_{\sun}$) than older clusters, for which the turnoff is
0.8 $M_{\sun}$. It is much more likely to find a donor with high-enough mass for
thermal-time-scale mass transfer to occur in a GC with a turnoff at
$\sim 1.1 M_{\sun}$. This can be easily understood by noting that all of the 
low-probability interactions that could produce such a binary in an old clusters 
are supplemented by a wealth of additional interactions involving stars of 
relatively high mass. Furthermore there is a clear prediction that thermal
mass transfer is much more likely in younger clusters. Calculations are underway
to quantify this effect.

\subsection{Conclusions}

M31 \gc s could have turned out to have X-ray properties almost
exactly like those of the Milky Way. In this case, studies of M31
would have told us more about the processes that go on in our
own Galaxy, simply because their larger numbers provide a
bigger arena for the relevant interactions to occur.
M31 \gc s could have turned out to have X-ray properties very 
different from  those of the Milky Way. In this case they
might have provided a gateway to our understanding of GCs in
external galaxies. 

The population of M31 GCs did turn out to be include clusters that
appear to have similar X-ray properties to those of the MW, as well
as a significant subset that have different X-ray properties.
Perhaps this is the best of both worlds, because we can now
hope to learn about processes occurring in our own GCs,
and also about processes that do not occur here or occur only
with very low probability.

For example, we might expect to find highly-luminous
thermal-time-scale mass transfer systems in galaxies with
populations of young clusters, and none such in galaxies such 
as our own, with only old populations of stars in GCs.
Thus, the relevance of the thermal-time-scale model we have
introduced, can be
tested by determining whether galaxies thought to include
more recently formed GCs  have more highly luminous GC X-ray sources.
Similarly, the importance of interactions with the host galaxy
can be tested by comparing GC \xrs s in galaxies that have more
massive bulges with GC \xrs s in galaxies with less massive bulges.

In short, {\it Chandra's} investigation of the M31 GC system has 
provided puzzles that  suggest many lines of investigation
for the further study of M31's GCs, GCs in external galaxies,
and theoretical work on binary formation and evolution in globular clusters.         
   
\begin{acknowledgements}
We gratefully acknowledge thoughful input from Anil Dosaj,
 Lars Hernquist,
 Margarita Karovska, Vicky Kalogera, Christopher Kochanek, Lucas Macri,
Jeff McClintock, and Andrea Prestwich.
This work was supported in part by NASA under
GO1-2091X; A.K.H.K. is supported by a Croucher Fellowship.  
\end{acknowledgements}

\end{document}